\documentclass[lettersize,journal]{IEEEtran}
\usepackage{amsmath,amsfonts}
\usepackage{algorithm}
\usepackage{array}
\usepackage[caption=false,font=normalsize,labelfont=sf,textfont=sf]{subfig}
\usepackage{textcomp}
\usepackage{stfloats}
\usepackage{url}
\usepackage{verbatim}
\usepackage{graphicx}
\usepackage{cite}
\usepackage{amsmath,amsfonts}
\usepackage{algorithmicx}
\usepackage{algpseudocode}
\usepackage{algorithm}
\usepackage{graphicx}
\usepackage{textcomp}
\usepackage{xcolor}
\usepackage{array}
\usepackage{xspace}
\usepackage{enumitem}
\usepackage{pifont}
\usepackage[colorlinks=true, linkcolor=black, urlcolor=black, citecolor=black]{hyperref}

\usepackage{amsmath,amsfonts,bm}
\usepackage{multirow}

\def\eqref#1{equation~\ref{#1}}

\def\1{\bm{1}}

\DeclareMathAlphabet{\mathsfit}{\encodingdefault}{\sfdefault}{m}{sl}
\SetMathAlphabet{\mathsfit}{bold}{\encodingdefault}{\sfdefault}{bx}{n}

\newcommand{\ie}{\textit{i}.\textit{e}.\xspace}

\usepackage{dsfont} 

\newcommand{\modelname}{CodeV\xspace}
\newcommand{\xname}{CodeV\xspace}

\usepackage{colortbl}
\usepackage{tabularray}
\UseTblrLibrary{booktabs}
\SetTblrInner[booktabs]{abovesep=0pt, belowsep=0pt, rowsep=0.5pt}
\SetTblrInner[booktabs]{cells = {cmd=\small}}

\NewTableCommand\seprule{\specialrule{\lightrulewidth,gray8}{2.5pt}{2.5pt}}
\NewTableCommand\uniquerule{\specialrule{\lightrulewidth,gray7,dashed}{2.5pt}{2.5pt}}
\definecolor{lightb}{RGB}{235,245,255}

\usepackage{graphicx}

\usepackage{pifont}%

\usepackage[nointegrals]{wasysym}
\usepackage{enumitem} %
\setlist[itemize]{leftmargin=*}

\usepackage{xspace}
\usepackage{xcolor}
\usepackage{mathtools}

\usepackage[capitalise, noabbrev]{cleveref}
\crefformat{section}{\S#2#1#3}
\Crefformat{section}{\S#2#1#3}

\usepackage{fontawesome}

\definecolor{codegreen}{rgb}{0,0.6,0}
\definecolor{codegray}{rgb}{0.5,0.5,0.5}
\definecolor{codepurple}{rgb}{0.58,0,0.82}
\definecolor{backcolour}{rgb}{0.95,0.95,0.92}
\definecolor{codeblue}{rgb}{0,0,0.7}

\makeatletter
\newcommand\codefontsize{\@setfontsize\codefontsize\@viiipt\@ixpt}
\makeatother

\usepackage{listings}
\lstdefinestyle{codestyle}{
    backgroundcolor=\color{backcolour},   
    commentstyle=\color{codegreen},
    keywordstyle=\color{codeblue},
    numberstyle=\tiny\color{codegray},
    stringstyle=\color{codepurple},
    basicstyle=\ttfamily\codefontsize,
    breakatwhitespace=false,         
    breaklines=false,                 
    captionpos=b,                    
    keepspaces=false,                 
    showspaces=false,                
    showstringspaces=false,
    showtabs=false,                  
    tabsize=2,
    numbers=none
}

\newcommand\Passat[1]{\mbox{Pass@{#1}}}

\algnewcommand{\LeftCommenta}[1]{\Statex \hspace{-1em} \(\triangleright\) #1}
\algnewcommand{\LeftCommentb}[1]{\Statex \hspace{1.3em} \(\triangleright\) #1}

\begin{document}

\title{\xname: Empowering LLMs with HDL Generation \\through Multi-Level Summarization}

\author{Yang Zhao,
        Di Huang,
        Chongxiao Li,
        Pengwei Jin,
        Muxin Song,
        Yinan Xu, 
        Ziyuan Nan,
        Mingju Gao, 
        Tianyun Ma,
        Lei Qi,
        Yansong Pan,
        Zhenxing Zhang,
        Rui Zhang,~\IEEEmembership{Member,~IEEE},
        Xishan Zhang,
        Zidong Du,~\IEEEmembership{Member,~IEEE},
        Qi Guo,~\IEEEmembership{Member,~IEEE},
        Xing Hu,~\IEEEmembership{Member,~IEEE}

\IEEEcompsocitemizethanks{

\IEEEcompsocthanksitem Yang Zhao, Chongxiao Li, Pengwei Jin, Ziyuan Nan, Mingju Gao, Yansong Pan are with the State Key Lab of Processors, Institute of Computing Technology, Chinese Academy of Sciences, Beijing, China, the University of Chinese Academy of Sciences, Beijing, China, and also with Cambricon Technologies.
\IEEEcompsocthanksitem Di Huang, Yinan Xu, Rui Zhang, and Qi Guo are with the State Key Lab of Processors, Institute of Computing Technology, Chinese Academy of Sciences, Beijing, China.
\IEEEcompsocthanksitem Muxin Song is with the School of Information Science and Technology, ShanghaiTech University, Shanghai, China.
\IEEEcompsocthanksitem Tianyun Ma, Lei Qi, Zhenxing Zhang are with the University of Science and Technology of China, Hefei, China, the State Key Lab of Processors, Institute of Computing Technology, Chinese Academy of Sciences, Beijing, China.
\IEEEcompsocthanksitem Zidong Du and Xing Hu are with the State Key Lab of Processors, Institute of Computing Technology, Chinese Academy of Sciences, Beijing, China, and Shanghai Innovation Center for Processor Technologies, Shanghai, China.
\IEEEcompsocthanksitem Xishan Zhang is with the State Key Lab of Processors, Institute of Computing Technology, Chinese Academy of Sciences, Beijing, China, and also with Cambricon Technologies.
}
}

\maketitle

\begin{abstract} 
The design flow of processors, particularly in hardware description languages (HDL) like Verilog and Chisel, is complex and costly. While recent advances in large language models (LLMs) have significantly improved coding tasks in software languages such as Python, their application in HDL generation remains limited due to the scarcity of high-quality HDL data. Traditional methods of adapting LLMs for hardware design rely on synthetic HDL datasets, which often suffer from low quality because even advanced LLMs like GPT perform poorly in the HDL domain. 
Moreover, these methods focus solely on chat tasks and the Verilog language, limiting their application scenarios. 

In this paper, we observe that: (1) HDL code collected from the real world is of higher quality than code generated by LLMs. 
(2) LLMs like GPT-3.5 excel in summarizing HDL code rather than generating it.
(3) An explicit language tag can help LLMs better adapt to the target language when there is insufficient data.
Based on these observations, we propose an efficient LLM fine-tuning pipeline for HDL generation that integrates a multi-level summarization data synthesis process with a novel Chat-FIM-Tag supervised fine-tuning method. The pipeline enhances the generation of HDL code from natural language descriptions and enables the handling of various tasks such as chat and infilling incomplete code.
Utilizing this pipeline, we introduce \modelname, a series of HDL generation LLMs. 
Among them, \modelname-All not only possesses a more diverse
range of language abilities, \ie Verilog and Chisel, and a broader scope of tasks, \ie Chat and fill-in-middle (FIM), but it
also achieves performance on VerilogEval that is comparable
to or even surpasses that of \modelname-Verilog fine-tuned on Verilog only, making them the first series of open-source LLMs designed for multi-scenario HDL generation.
\end{abstract}

\begin{figure*}[h]
    \centering
    \includegraphics[width=0.85\linewidth]{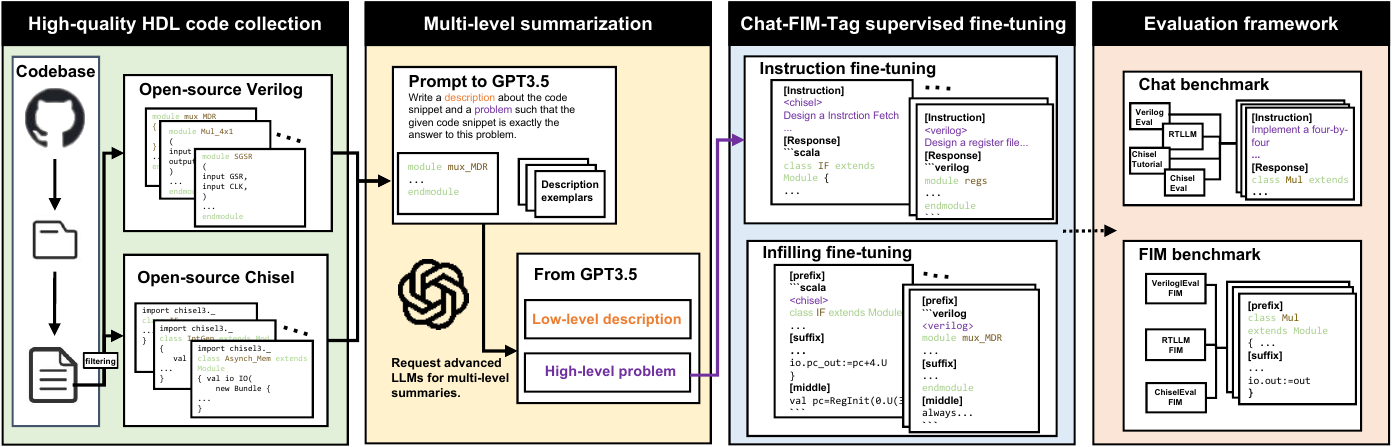}
    \caption{\textbf{The \xname framework overview.} We first collect and filter high-quality HDL modules from open-source codebases. The modules are then sent to GPT-3.5 to request multi-level summaries. Pairing high-level descriptions with corresponding modules, the high-quality dataset is utilized to fine-tune base LLMs, yielding \xname models.}
    \label{fig:overview}
\end{figure*}
\vspace{3pt}
\section{Introduction}
\IEEEPARstart{T}{he} design flow of processors is expensive and complex, especially the coding of the hardware description language (HDL) like Verilog and Chisel~\cite{bergeron2003writing, foster2022wilson, kabylkas2023effective}.
Recent advances in large language models (LLMs) have not only delivered impressive performance in coding tasks for software languages such as Python~\cite{instructgpt,gpt4-report, wizardcoder, magicoder, wavecoder, octopack, opencodeinterpreter} but have also unveiled promising avenues for the direct generation of HDL from natural language descriptions.
However, the good performance of LLMs on software coding tasks is attributed to the large amount of high-quality description-code data available in the software field, whereas in the field of hardware design, high-quality description-code HDL data is very scarce.
For example, in the largest open-source code dataset, The Stack v2~\cite{starcoder2}, the Scala data accounts for only 10.1M (with Chisel constituting only a small part of the Scala language), the Verilog data is merely 1.91M, whereas the Python data amounts to 80.6M – a difference of several tens of times.

Although initial efforts have explored adapting LLMs for hardware design by fine-tuning them on synthesized Verilog data~\cite{liu2023rtlcoder, pei2024betterv, gao2024autovcodersystematicframeworkautomated, liu2025craftrtlhighqualitysyntheticdata, yang2025haven, cui2024origen}, these approaches have not yet effectively addressed this problem. This shortfall primarily stems from current data synthesis methods, which heavily depend on advanced LLMs—such as GPT—to generate both descriptions and corresponding code. However, because even these advanced LLMs perform poorly in the HDL domain, the quality of the synthesized data remains low. For instance, while GPT-4 achieves an 88.4\% pass@1 score on the Python benchmark HumanEval~\cite{codex, evalplus}, it only reaches 43.5\% on the VerilogEval-Human benchmark~\cite{verilogeval}. Moreover, existing methods focus solely on chat tasks and the Verilog language—that is, they generate Verilog code from natural language descriptions—and are not equipped to handle additional scenarios, such as infilling incomplete code, or other languages, such as Chisel.

\begin{figure}[t]
    \begin{center}
    \includegraphics[width=0.48\textwidth]{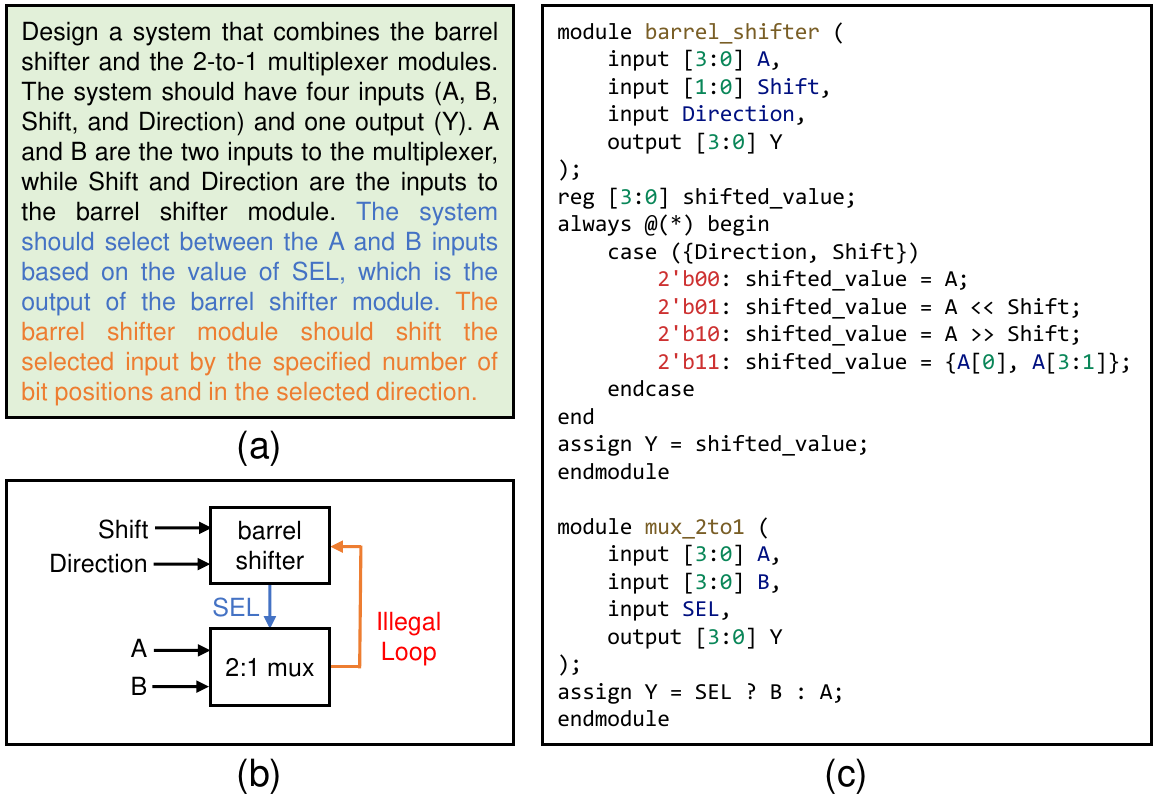}
    \end{center}
    \caption{GPT-generated Verilog dataset contains unrealistic code. (a) A synthetic description obtained from RTLCoder~\cite{liu2023rtlcoder}, which is unrealistic. (b) The circuit's block diagram according to the description, which contains illegal loops in the circuit. (c) The corresponding incorrect Verilog code in the dataset.}
    \label{fig:diss_rtlcoder}
\end{figure}

Regarding these challenges, we have observed the following:
\textbf{(1) HDL code collected from the real world is of higher quality than that generated by LLMs, and LLMs excel in summarizing HDL code rather than generating it.}
HDL code collected from the real world has fewer syntax errors, clearer semantics, and better conforms to the code distribution in practical applications compared to HDL code generated by LLMs.
Figure~\ref{fig:diss_rtlcoder}a-c illustrates an example observed from the GPT-generated HDL dataset from RTLCoder~\cite{liu2023rtlcoder}. Figure~\ref{fig:diss_rtlcoder}a shows a description obtained by combining two circuit descriptions. We draw a block diagram based on this description, and as shown in Figure~\ref{fig:diss_rtlcoder}b, this description leads to a loop in a combinational circuit, which is illegal. Figure~\ref{fig:diss_rtlcoder}c depicts the circuit generated by GPT-3.5 based on this incorrect description, which fails to integrate the shifter and multiplexer, and the internal implementation of the shifter module is also incorrect.
\textbf{(2) LLMs excel in summarizing HDL code rather than generating it.
}
As shown in Table~\ref{tab:summarize_vs_generate}, where we manually assess whether the descriptions produced by GPT-3.5 accurately correspond to the Verilog code, although GPT-3.5's poor performance in generating Verilog ($33.5\%$ pass@1), it indeed offers the ability to describe the functionality of given Verilog code ($64.7\%$ pass@1).
\textbf{(3) An explicit language tag can help LLMs better adapt to the target language when there is insufficient data.
} For example, when generating both Verilog and Chisel with even scarcer data, using the language tags ``\texttt{<Verilog>}'' and ``\texttt{<Chisel>}'' to differentiate between them can simultaneously enhance the model’s performance in both languages. A detailed supporting experiment is shown in Table~\ref{tab:ablation_chisel_verilog}.

Based on these observations, we propose an efficient LLM fine-tuning pipeline that targets HDL generation.
As shown in Figure~\ref{fig:overview}, this pipeline includes a data synthesis process and a training process.
For the data synthesis process, we employ \textbf{multi-level summarization} for generating the high-quality dataset from raw code.
Instead of synthesizing descriptions first and then getting the corresponding code from an advanced LLM, we prompt the LLM with HDL code and let the LLM generate the corresponding natural language descriptions.
For the fine-tuning process, to more closely align with real-world design workflows, we use the Chat-FIM (\ie fill-in-middle) hybrid fine-tuning method with a language tag that enables LLMs to effectively handle different languages even when limited data is available.

\begin{table}[t]
\caption{
Summarizing Verilog V.S. generating Verilog. 
}
\centering
\begin{tabular}{cc}
\toprule
   Task               & \Passat{1} (\%) \\
\midrule
Verilog Summarization       & 64.7                     \\
Verilog Generation         & 33.5                     \\
\bottomrule
\end{tabular}
\vspace{5pt} %

\label{tab:summarize_vs_generate}
\end{table}

Building upon this pipeline, we introduce \modelname, a series of instruction-tuned HDL generation LLMs.
Specifically, in the data synthesis process, we first crawl over 1.5M Verilog modules and 24K Chisel modules from GitHub, filter them, and obtain a deduplicated, self-contained, and syntax-correct set containing 165K high-quality Verilog modules and 18.7K Chisel modules. Based on these HDL modules, we use GPT-3.5 to generate fine-grained functional descriptions that more accurately capture the functionality of HDL, although there remains a gap with real-world use cases. We then let GPT-3.5 summarize high-level function specifications, which are closer to real-world scenarios, thereby obtaining a high-quality dataset of description-code pairs. 
In the fine-tuning process, we employ the Chat-FIM-Tag supervised fine-tuning method, which enables \modelname to handle both Chat (generating code based on natural language descriptions) and FIM (infilling incomplete code) tasks and two different HDL, Verilog and Chisel.

Additionally, to evaluate the generation and FIM ability of LLMs on Chisel, we propose two Chisel benchmarks, ChiselTutorial and ChiselEval, and extend existing Verilog benchmarks to an FIM format. 
Among them, ChiselTutorial is a collection of introductory Chisel problems, and ChiselEval is the Chisel version of the VerilogEval translation.
Experimental results show that 
\modelname-All not only possesses a more diverse range of language abilities and a broader scope of tasks, but it also achieves performance on VerilogEval that is comparable to or even surpasses that of \modelname-Verilog, making them the first series of open-source LLMs designed for multi-scenario HDL generation.
Specifically, \modelname-All-QC achieves state-of-the-art (SOTA) results within open-source LLMs on both VerilogEval~\cite{verilogeval} and RTLLM~\cite{rtllm}, and surpasses the ever-reported SOTA LLM CraftRTL~\cite{liu2025craftrtlhighqualitysyntheticdata} by 4.1\% pass@1 and the commercial LLM GPT-3.5 by 35.2\% pass@1 on VerilogEval-Machine.

The major contributions of this paper are as follows:
\begin{itemize}
    \item We propose an efficient LLM fine-tuning pipeline that targets HDL generation, including the multi-level summarization data synthesis process and Chat-FIM-Tag supervised fine-tuning process. This pipeline enables the generation of HDL for general scenarios.
    \item Building on this method, we introduce a series of state-of-the-art (SOTA), open-source Verilog generation LLMs, named \modelname-Verilog. Notably, \modelname-Verilog-QC achieves an 80.1\% pass@1 on the VerilogEval-Machine benchmark and 59.2\% on the VerilogEval-Human benchmark—surpassing GPT-3.5, GPT-4. It also attains a 96.6\% syntax pass rate and a 51.7\% function pass rate on the RTLLM benchmark, outperforming RTLCoder, the previous SOTA open-source LLM.
    \item We propose two Chisel benchmarks, ChiselTutorial and ChiselEval, and extend them along with existing Verilog benchmarks to an FIM format. These benchmarks provide a more comprehensive evaluation process for LLM-based HDL generation.
    \item To show the generalizability of our pipeline, we introduce \modelname-All, the first open-source LLM series designed for multi-scenario HDL generation. It demonstrates remarkable performance across different HDL languages (Verilog and Chisel) and multiple tasks (Chat and FIM), showcasing the generalization capability of our method.
    \item We plan to open-source our dataset, which contains 185K high-quality description-code pairs (165.3K for Verilog and 18.7K for Chisel), to foster advancement and collaboration within the LLM, electronic design automation (EDA), and programming language communities.
\end{itemize}

\section{Related Work}

\begin{figure*}[h!]
    \centering
    \includegraphics[width=0.9\linewidth]{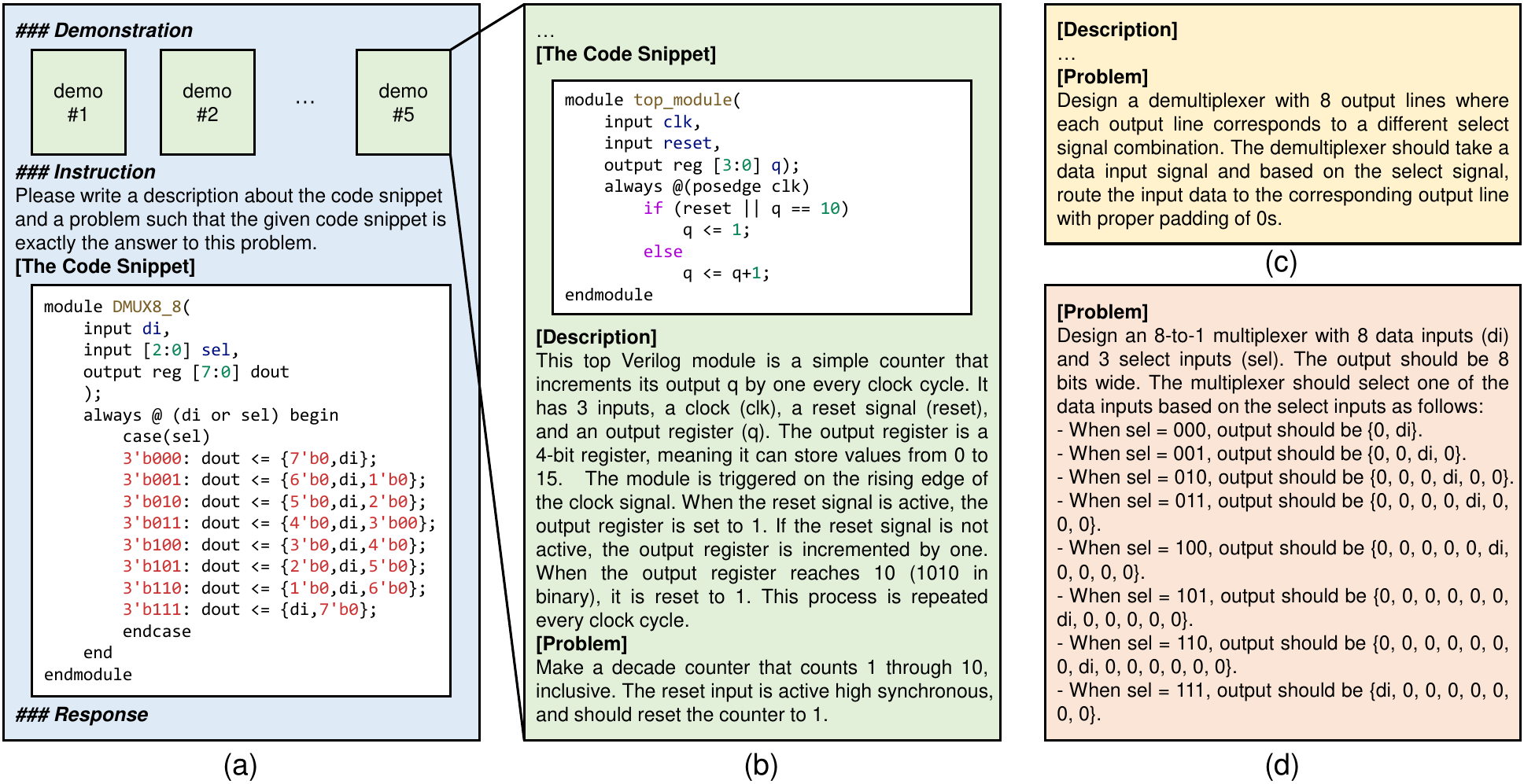}
    \caption{An actual example of the prompt for multi-level summarization. (a) The prompt provided to GPT-3.5. (b) An example of the demonstrations, with code, low-level descriptions, and high-level summaries. (c) Summaries responded from GPT-3.5 with and (d) without multi-level summarization.}
    \label{fig:multi_level_sum_demo}
\end{figure*}

\subsection{LLMs for HDL Generation}
Thakur et al.~\cite{thakur2023benchmarking} collected approximately 400MB of Verilog files from GitHub and textbooks to fine-tune CodeGen-16B~\cite{codegen}. However, the dataset was not refined, resulting in a blend of text and code, with some Verilog code potentially containing syntax errors.
ChipNemo~\cite{liu2023chipnemo} used LLaMA2~\cite{touvron2023llama} as its base model, performing domain-adaptive pre-training on public datasets and NVIDIA's internal chip design files (including HDL code, EDA scripts, and configuration files, etc.), and fine-tuning on public instruction datasets and an expert-crafted domain dataset with 1430 examples.
However, likely due to the dataset not being refined for Verilog generation tasks, the 70B version model of ChipNemo even underperforms its base model on the VerilogEval-Human benchmark.
RTLCoder~\cite{liu2023rtlcoder} uses GPT to synthesize instruction-code pairs, starting by generating instructions based on a keywords pool and a source code pool, then mutating these instructions to increase their number and diversity, and finally having GPT generate Verilog code based on these synthesized instructions to form a fine-tuning dataset. This approach may lead to unrealistic instructions, and the code quality is limited by GPT's generation ability. 
BetterV~\cite{pei2024betterv} collects Verilog code from GitHub and expands its dataset by translating Verilog code into C to create C-Verilog pairs, leveraging the base model's knowledge in C to enhance Verilog generation capabilities. 
AutoVCoder~\cite{gao2024autovcodersystematicframeworkautomated} collects data from up to 20,000 GitHub repositories based on its proposed automated dataset generation method, and generates corresponding question-code pairs through LLM, using different keywords to increase code diversity.
OriGen~\cite{cui2024origen} collected open-source RTL code samples and generated detailed descriptions of the corresponding code using Claude3-Haiku. It then regenerated RTL code based on the detailed descriptions to build a code enhancement dataset, and proposed an error correction dataset by modifying parts of the main content of the code.
CraftRTL~\cite{liu2025craftrtlhighqualitysyntheticdata} identified the difficulties of non-text representations in synthetic data generation and the variability in performance during benchmark training. It proposed a targeted fine-tuning dataset for LLMs, which faced challenges in solving Karnaugh maps, state transition diagrams, and waveform issues in Verilog code problems, and significantly alleviated the model's performance on related issues.
HAVEN~\cite{yang2025haven} addresses issues such as symbolic illusion, knowledge illusion, and logical illusion. Based on the proposed illusion classification method, it constructs knowledge-enhanced and logic-enhanced datasets, and proposes a symbolic understanding chain method that integrates chain-of-thought (CoT).
In parallel with our work, MG-Verilog~\cite{zhang2024mg} uses the summarization technique to create a multi-grained description-Verilog dataset. However, our focus is on different languages (\ie Verilog and Chisel) and scenarios (\ie Chat and FIM). While MG-Verilog aims to generate Verilog code from varying levels of descriptions, leading them to modify the original VerilogEval benchmark to fit their needs, our goal is to enhance the overall performance of HDL generation models utilizing the mismatch between generation and summarization ability and demonstrate their potential in practical module design.

In the field of hardware circuit design, traditional low-level hardware description code design related to HDL such as Verilog is difficult to meet the requirements for code readability, modularity, and scalability. The Chisel~\cite{chisel}, which adopts an object-oriented approach to circuit design, based on the Scala language has emerged.
ChatChisel~\cite{chatchisel} implemented the design of an RV32I RISC-V CPU with the help of GPT-3.5-Turbo, achieving a 31.86\% performance improvement over Verilog-based hardware design levels. However, the data scarcity challenge of Chisel is even more severe than that of Verilog, which leads to poor performance on fine-tuning. Moreover, there is a lack of relevant benchmarks to measure the hardware circuit generation capabilities of LLMs on Chisel.

\subsection{Frameworks on HDL Generation}
Instead of training specific models, some work focuses on designing generation frameworks for HDL tasks to improve general-purpose LLMs' performance. Chip-Chat~\cite{Blocklove_2023} and ChipGPT~\cite{chang2023chipgpt} use multi-round EDA and human feedback to assist LLMs for Verilog generation. Lu et al.~\cite{rtllm} introduced a self-planning method to improve generation accuracy, where LLMs plan before generating Verilog.
OriGen~\cite{cui2024origen} uses feedback loop to generate Verilog and incorporates a teacher-student learning framework to correct the dataset in order to help the model better understand error messages.
AutoVCoder~\cite{gao2024autovcodersystematicframeworkautomated} introduces retrieval-augmented generation (RAG) technology into Verilog design problems and designs a framework containing an example retriever and a knowledge retriever.
RTLFixer~\cite{tsai2024rtlfixer} combines RAG and ReAct prompts to automatically fix syntax errors in Verilog code using large language models.
Andre Nakkab et al.~\cite{RomewasNotBuiltinaSingleStep} proposed a hierarchical prompting technique that uses a layered approach to automatically generate complex hardware modules and completed a processor entirely designed by LLM.
VerilogCoder~\cite{ho2025verilogcoderautonomousverilogcoding} proposed a system composed of multiple agents that can autonomously write Verilog and use Verilog tools to fix syntax and functional errors.
AIvril~\cite{islam2024aivrilaidrivenrtlgeneration} designed a framework that combines a Code Agent and a Review Agent to write Verilog code and testbench, and to perform automatic self-correction and functional verification.
RTL Agent~\cite{RTLAgent} explored high-quality Verilog generation tasks by designing a Reflexion workflow generated by Verilog.
These methods are orthogonal to our work, which forms the foundation for them, as they all require a code model capable of generating HDL.

In contrast to these methods, CodeV employs a novel dataset construction approach with a validated insight into the difficulty of code generation and code summarization, utilizing multi-level summarization to generate high-quality instruction tuning datasets from high-quality Verilog code, achieving SOTA results.

\section{Methods}
\subsection{High-quality HDL Code Collection}
\label{sec:verilog_collection}

\paragraph{Verilog code collection}
High-quality real-world Verilog code is the foundation of our approach. 
We first crawl Verilog and SystemVerilog files from GitHub, ensuring each file contains a complete module, i.e. contains at least one ``\texttt{module}'' and ``\texttt{endmodule}'' pair. 
Files with external references, \ie code containing ``\texttt{include}'' or ``\texttt{import}'' keywords, are excluded.
This is because reconstructing these reference relationships from the crawled files is challenging, and we do not want LLMs to generate these non-self-contained Verilog codes after training. 
Then we deal with the comments interspersed throughout the files.
These comments can be categorized into two types: those related to Verilog implementation, such as signal descriptions, module explanations, test cases, etc.; and those unrelated to implementation, such as licenses, company and author information, logs, etc.
The former is essential for LLMs to understand the code and should be kept, while the latter should be avoided.
Therefore, we use regular expressions to delete these implementation-unrelated comments.
Next, files exceeding $4096$ characters are deleted, as longer files would be truncated during training, resulting in incomplete file content and degrading the dataset quality.

\paragraph{Chisel code collection}
In order to enable our model to have multi-lingual capabilities, we create a Chisel dataset. Similar to the method used for creating the Verilog dataset, we obtain 1,651 repository projects related to Chisel currently on GitHub through the GitHub API. 
We extract Chisel code from these projects. We retain files with the ``.scala'' extension that import the Chisel package in their content. These characteristics distinguish hardware design code written in the Chisel language from others.

\paragraph{Data deduplication}
This process aims to filter out similar data in the training dataset.
Both Verilog and Chisel employ the same deduplication method.
Following \cite{thakur2023benchmarking}, we also use MinHash and Jaccard similarity to filter out duplicate code. 
Each Verilog file is hashed into a 128-dimensional vector using the MinHash API provided by datasketch~\cite{eric_zhu_2024_11462182}.
We go through the files in order, and if the Jaccard similarity between a Verilog file and any preceding file exceeds a certain threshold, the file is considered a duplicate and filtered out.
In our processing, this threshold is set to 0.8.
\paragraph{Quality enhancing} To further enhance the quality of the dataset, we compile each file with Icarus Verilog~\cite{iverilog} for automatic syntax checking.
Files that fail to compile are removed.
\begin{figure}[t]
    \centering
    \includegraphics[width=1\linewidth]{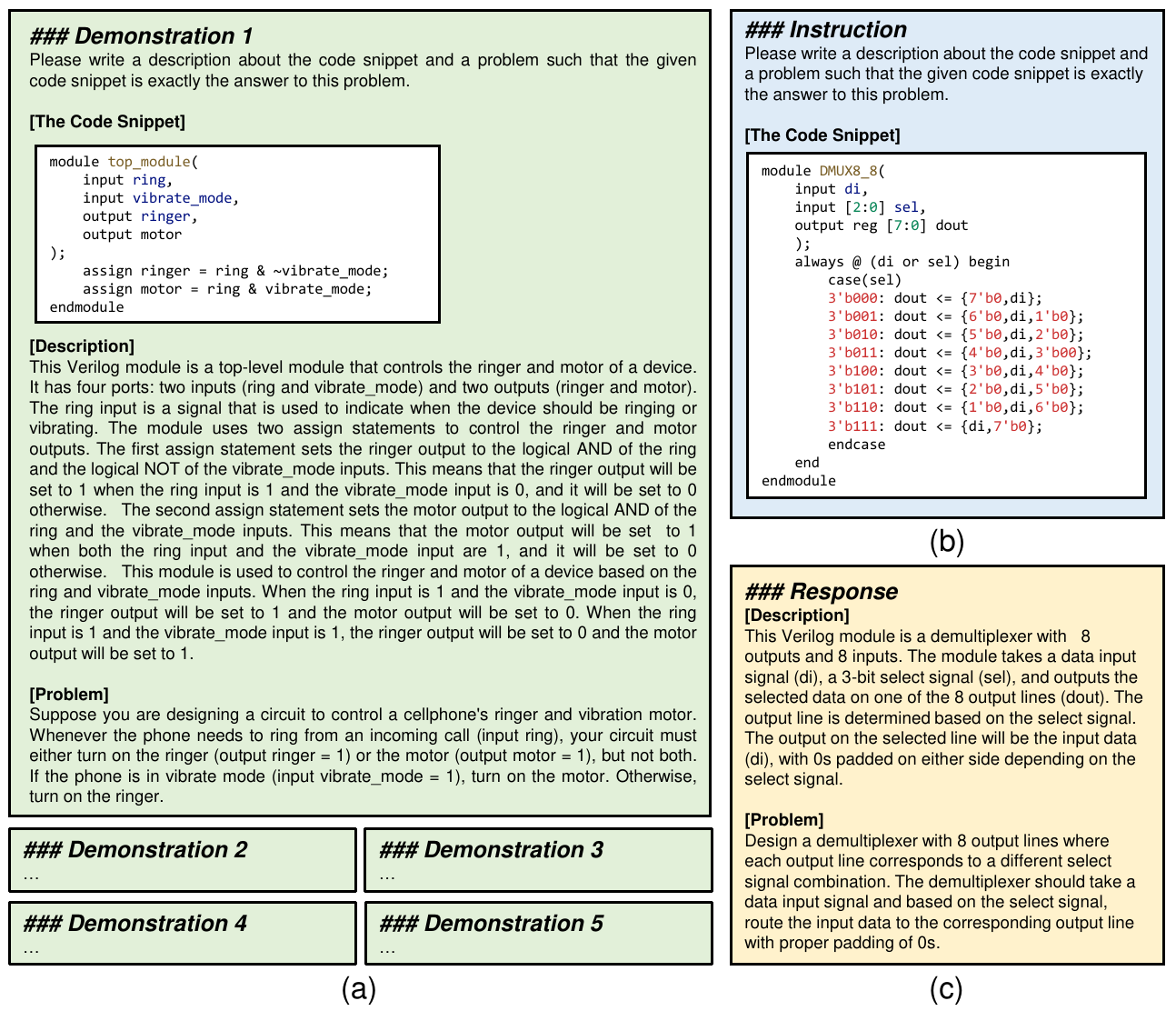}
    \caption{A detailed example showing all prompts used for multi-level summarization with Verilog dataset. (a) Demonstrations given to GPT-3.5. (b) A Verilog data snippet in CodeV generating corresponding Instructions, which can be adjusted by modifying the Code Snippet. (c) The response from GPT-3.5, consisting of the Description and Problem sections.}
    \label{fig:apps}
\end{figure}

\subsection{Multi-level Code Summarization}
\label{sec:multi_level_sum}

As mentioned before, manual annotation is prohibitively time-consuming and costly. 
Hence, we use GPT-3.5 to generate high-level summaries for each Verilog and Chisel module as its design specification. As analyzed in VerilogEval~\cite{verilogeval}, when required for summarizing, LLMs often produce verbose descriptions, preferring line-by-line explanations over high-level summaries. To address this issue, we introduce a multi-level summarization method, employing few-shot learning~\cite{brown2020language} to guide GPT-3.5 in first producing detailed descriptions and then abstracting high-level summaries. 

Figure~\ref{fig:multi_level_sum_demo} presents a Verilog example from our actual process, involving the generation of requirement descriptions for a demultiplexer module collected from GitHub. Figure~\ref{fig:multi_level_sum_demo}a illustrates our prompt, which includes $n$ examples with code snippets, detailed descriptions, and problem summaries, followed by the demultiplexer code, and a request for GPT to imitate the examples by first providing a detailed description and then a problem summary. Figure~\ref{fig:multi_level_sum_demo}b illustrates an example used for GPT's few-shot learning. Figure~\ref{fig:multi_level_sum_demo}c shows GPT's response using multi-level summarization, while Figure~\ref{fig:multi_level_sum_demo}d shows the response without multi-level summarization by deleting the detailed description in the demonstrations. We observe that the description in Figure~\ref{fig:multi_level_sum_demo}c is more concise and aligns with actual Verilog design specifications. In contrast, the description in Figure~\ref{fig:multi_level_sum_demo}d is verbose and even inaccurately counts the number of zeros in output signals, leading to a contradictory explanation.

The prompt details are shown in Figure~\ref{fig:apps} where we add more details about one of the 5 demonstrations. The 5 demonstrations are ``ringer'' (fully displayed), ``dff16e'', ``count1to10'', ``lfsr5'', and ``gatesv100''.

\begin{figure}
    \vspace{1mm}
    \begin{center}
    \includegraphics[width=0.35\textwidth]{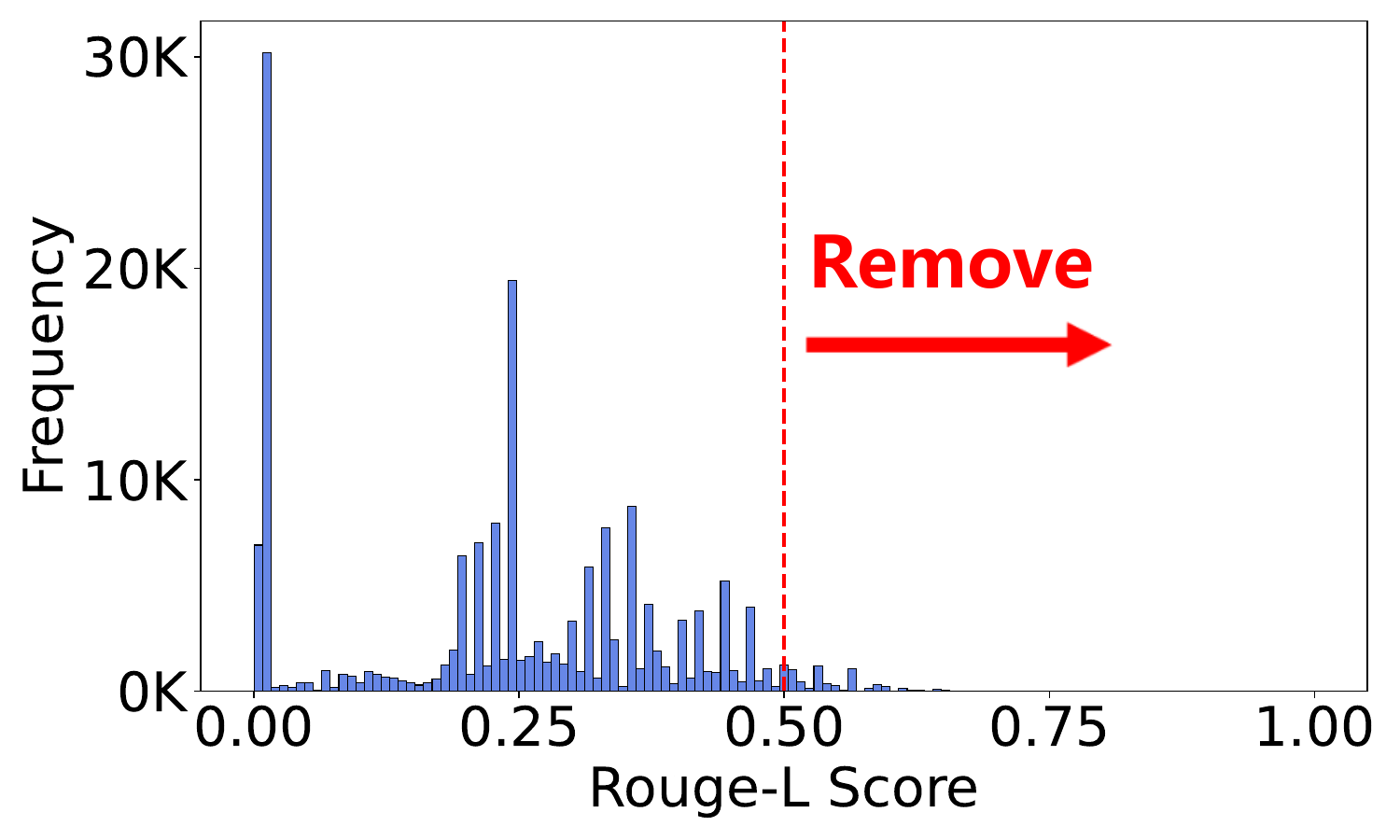}
        \caption{\small{The similarity distribution of our dataset against benchmark data. Data with Rouge-L scores greater than $0.5$ are removed for decontamination.}}\label{fig:rrrouge-l}
    \end{center}
    \vspace{-10px}
\end{figure}

\subsection{Data Decontamination}

To ensure a fair evaluation, we conduct decontamination for both Verilog and Chisel training data to remove sequences similar to benchmarks (details about our Chisel benchmark is shown in Section~\ref{sec:evaluation framework}). Specifically, we use Rouge-L~\cite{lin-2004-rouge}, a widely used similarity metric, to measure the similarity between data:
\begin{equation}
    ROUGE-L(C_{train}^i) = \max\limits_{j}(\frac{(1+\beta^2)LCS(C_{train}^i, C_{test}^j)}{len(C_{train}^i)+\beta^2 len(C_{test}^j)}),
\end{equation}
where $C_{train}^i$ denotes the i-th Verilog code from the training data, $C_{test}^j$ denotes the j-th Verilog code in the test data, and $LCS(C_{train}^i, C_{test}^j)$ denotes the length of a longest common subsequence of $C_{train}^i$ and $C_{test}^j$.
$ROUGE-L(C_{train}^i) \in [0, 1]$, with higher values indicating greater similarity. Following RTLCoder~\cite{liu2023rtlcoder}, we set $\beta=1.0$ and remove data from the training set that have a Rouge-L score greater than $0.5$ with any of the benchmarks, thereby preventing test data leakage. The Rouge-L plot is shown in Figure~\ref{fig:rrrouge-l}.
Through the above-mentioned effort, we obtain high-quality real-world Verilog code. Next, we will introduce the fine-tuning method.

\subsection{Chat-FIM-Tag supervised fine-tuning}
We introduce our Chat-FIM-Tag supervised fine-tuning method to alleviate the challenges of multi-scenario chat and infilling tasks, and that different languages may affect each other's performance when the training data is scarce.

Due to the inherent contextual dependencies in programming languages, relying solely on the next-token prediction of LLMs is insufficient to achieve satisfactory FIM performance. To enhance the FIM capability of the model, we conduct infilling fine-tuning following \cite{bavarian2022efficient}. Specifically, we randomly partition each training document into prefix, middle, and suffix sections, then concatenate them using special FIM tokens. Both line-level and character-level spans are employed for the partition, maintaining a 2:1 ratio between them. 
There are two distinct infilling modes for FIM method: PSM (Prefix-Suffix-Middle) and SPM (Suffix-Prefix-Middle). The PSM mode concatenates sections in the order of prefix, suffix, and middle, which is particularly suited for scenarios where both the prefix and suffix are non-empty on either side of the middle section. Conversely, the SPM mode follows the sequence of suffix, prefix, and middle for concatenation.
According to \cite{deepseekcoder,hui2024qwen2}, we employ the PSM (Prefix-Suffix-Middle) mode for infilling fine-tuning. Finally, we have processed the training document into the following tokenized version:
\begin{equation*}
    \texttt{<PRE>\{prefix\}<SUF>\{suffix\}<MID>\{middle\}<EOT>}
\end{equation*}
where ``\texttt{<PRE>}'', ``\texttt{<MID>}'', ``\texttt{<SUF>}'', and ``\texttt{<EOT>}'' need to be modified according to the model's tokenization. It is noteworthy that we preserve the loss for all the prefix, middle, and suffix sections in this study.

When the data for a single language is limited, including multiple languages can affect the performance of that language. To address this challenge, we have empirically found that emphasizing the distinctions between languages during training helps improve the performance of each language. Specifically, we incorporate language-specific tags, that is ``\texttt{<Verilog>}'' and ``\texttt{<Chisel>}'', during training.

\begin{figure}[t]
    \begin{center}
        \includegraphics[width=0.48\textwidth]{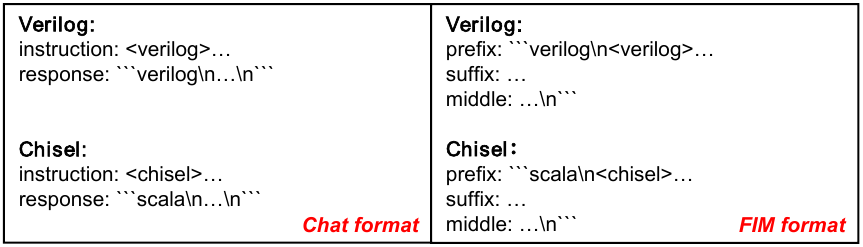}
    \end{center}
    \caption{The training data format. Here, ``\texttt{<verilog>}" and ``\texttt{<chisel>}" are used to identify the language of the task, while ``\texttt{```verilog\textbackslash n...\textbackslash n```}" and ``\texttt{```scala\textbackslash n...\textbackslash n```}" are used to mark the scope of code segments.}
    \label{fig:training_format}
\end{figure}
\begin{table}[!]
\caption{Composition of the training datasets.
}
\centering
\setlength{\tabcolsep}{1mm}
\footnotesize
\small
\label{tab:dataset_cmp}

\begin{tabular}{cccc}
\toprule
   Dataset& Verilog (k)& Chisel (k)&FIM rate (\%)\\
\midrule
165k-Verilog& 165.3& 0&0\\
184k-Verilog-Chisel& 165.3& 18.7&33.3\\
\bottomrule
\end{tabular}

\end{table}
Based on these, we construct two training datasets, as presented in Table~\ref{tab:dataset_cmp}.
To enable models to better distinguish between different language tasks and identify the start and end positions of code, we add tags to the code in the 184k-Verilog-Chisel dataset before the tokenization process. The format can refer to Figure~\ref{fig:training_format}. 

Finally, we fine-tune base models on our datasets. Given a natural language description, the LLM with parameter $\theta$ generates code repeatedly drawing from $p_{\theta}(x_t|x_{<t})$ and using $x_t$ as part of the input for the subsequent prediction. 
Following prior work, the LLM is designed to reduce the negative log-likelihood across the dataset D $\{x^1, ..., x^{|D|}\}$~\cite{keskar2019ctrl}:
\begin{equation}
    \mathcal{L} = -\frac{1}{|D|}\sum_{i=1}^{|D|}\frac{1}{n}\sum_{t = 1}^{n}\log p_{\theta}(x^i_t|x^i_{<t}).
\end{equation}

\section{Evaluation framework}
\label{sec:evaluation framework}
In this section, we introduce our benchmarks based on the proposed Chat and FIM testing tasks. In the first part, we modify the open-source Chisel tutorial to design the ChiselTutorial benchmark for Chisel evaluation. Additionally, we adapt the benchmark problems from VerilogEval, considering the diverse hardware programming challenges it covers, to create ChiselEval for Chisel evaluation.  In the second part, we reformat three Verilog and Chisel benchmarks, VerilogEval, RTLLM, and ChatEval, to establish FIM benchmarks for FIM evaluation, namely VerilogEval-FIM, RTLLM-FIM, and ChatEval-FIM. Details are introduced as follows.

\begin{figure}[t]
    \begin{center}
    \includegraphics[width=0.48\textwidth]{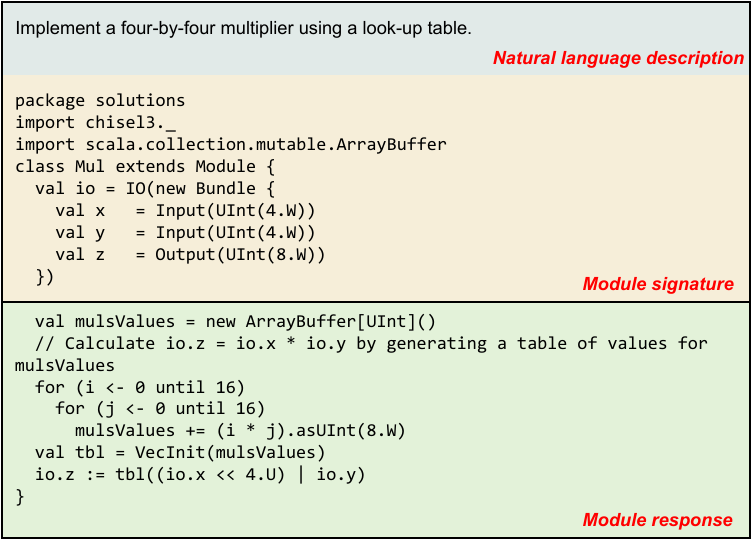}

    \end{center}
    \caption{An example from ChiselTutorial.}
    \label{fig:ChiselTutorial}
\end{figure}

\subsection{Chat Benchmark}
\label{exp:Chat_benchmark}
\paragraph{ChiselTutorial Benchmark}
We find that the current LLMs still have a lot of syntax errors in Chisel programming capabilities. Therefore, it is necessary to conduct the most basic Chisel introductory tests. 
We modify ChiselTutorial\footnote{\url{https://github.com/ucb-bar/chisel-tutorial}}, an official programming exercise of Chisel, to form our ChiselTutorial benchmark which evaluates the basic Chisel capabilities of LLMs. ChiselTutorial is an official example code presented for beginners learning Chisel, aiming to demonstrate the latest Chisel programming syntax to Chisel novices. This example includes some relatively simple Chisel programming tasks, covering basic hardware design problems from multiplexers and arithmetic operations to counters.
Specifically, we extracted 17 programming problems from the ChiselTutorial as evaluation problems in our benchmark. 
We used the original code as the answers and then extracted and modified the comments that introduce the problems in the original code as the natural language descriptions. After each problem description, we added the module header definitions from the original code file to ensure that the generated Chisel code can be accurately called by the testbench.
For evaluation, we use the official testbench from ChiselTutorial and run the code using the Scala build tool set. We provide reference correct code and include inputs that can be processed by LLMs along with the corresponding module headers.
An example problem of ChiselTutorial is shown in Figure~\ref{fig:ChiselTutorial}.

\paragraph{ChiselEval Benchmark}
To further evaluate the Chisel programming ability of LLMs, we design the ChiselEval benchmark. Referring to the problem structure in VerilogEval, we create ChiselEval by rewriting the answers to all 156 problems in VerilogEval-Human to Chisel.
Similar to VerilogEval, ChiselEval also extensively covers fundamental problems in hardware circuit design, such as basic signal processing, multiplexer design, and so on, ensuring that our designed benchmarks align more closely with actual design problems in Chisel. 
We modify problems in VerilogEval to make them suitable for Chisel generation. When adapting the natural language specifications in the problems, we ensure that the circuit functionality and logical implementation remain consistent with the original VerilogEval benchmark. We retain and adapt problems involving graphs or tables in VerilogEval, such as those with Karnaugh maps or truth tables, to evaluate the model's ability to design circuits based on these non-textual contents.
For each design, our benchmark also provides the corresponding verification environment. 
When all test cases match the expected output in the top module, it indicates that the LLM has correctly generated the code.

\begin{figure}[t]
    \begin{center}
    \includegraphics[width=0.48\textwidth]{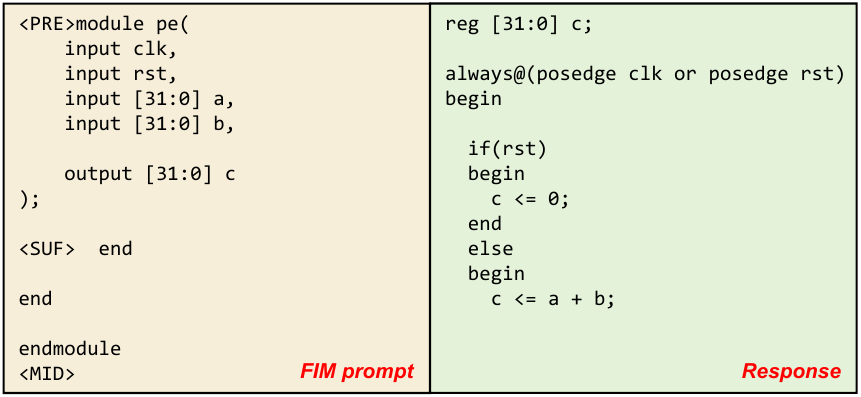}
    \end{center}
    \caption{An example of a VerilogEval-FIM testing framework.}
    \label{fig:FIM_testformat}
\end{figure}

\subsection{FIM Benchmark}
\label{exp:FIM_benchmark}
We introduce three FIM benchmarks: VerilogEval-FIM, RTLLM-FIM, and ChiselEval-FIM, which are constructed by adapting VerilogEval\cite{verilogeval}, RTLLM\cite{rtllm}, and ChiselEval. These benchmarks are specifically designed to evaluate the FIM capabilities of models in the Verilog and Chisel. Following \cite{incoder,bavarian2022efficient}, we implement single-line, multi-line, and random-span infilling for each FIM benchmark.

In the case of single-line infilling, we mask one non-empty line of the canonical code implementation to generate the FIM task. Multi-line infilling adopts a similar approach, with the distinction that the masked region may contain multiple non-empty lines.
With random-span infilling, we randomly mask a non-empty middle span at the character level. 
From each canonical code implementation, we randomly select one instance from each of the three generated task types to serve as the final FIM (Fill-in-Middle) task, ensuring each implementation yields exactly one FIM task for each infilling type while maintaining uniformity across all types.

For the three benchmarks, we preserve the module headers from their canonical code implementations to ensure testing reliability, avoiding potential issues caused by variations in generated module header names. For each circuit across the benchmarks, we establish tasks corresponding to all three infilling types. Consequently, the VerilogEval-FIM Benchmark consists of two parts: VerilogEval-FIM-Machine and VerilogEval-FIM-HUMAN, containing 429 and 468 tasks respectively. RTLLM-FIM comprises a total of 87 tasks, and the ChiselEval-FIM Benchmark contains 468 tasks. The distribution among the three infilling types is balanced, with equal numbers for each type.

Taking the VerilogEval-FIM Benchmark as an example, the test format is illustrated in Figure~\ref{fig:FIM_testformat}. In this format, the tokens ``\texttt{<PRE>}", ``\texttt{<MID>}", and ``\texttt{<SUF>}" need to be modified according to the model's specific FIM tokens. Through these three FIM benchmarks, the model's code-infilling capabilities in both Verilog and Chisel hardware programming languages can be effectively evaluated.

\section{Experiments}
\label{sec:experiments}

We conduct comprehensive experiments to investigate that:
\begin{itemize}
    \item How does \xname-Verilog, LLMs fine-tuned only with Verilog data, perform on Verilog across multiple benchmarks and settings (Section~\ref{exp:main})?
    \item How does \xname-All, fine-tuned with Verilog and Chisel data, perform on multi-lingual and multi-scenario tasks (Section~\ref{exp:codev-chisel-fim})?
    \item The effectiveness of multi-level summarization and Chat-FIM-Tag supervised fine-tuning (Section~\ref{exp:ablation}).
    \item Further analysis on the impact of dataset size, and sampling temperature (Section~\ref{exp:further_analysis}).
\end{itemize}

\begin{table*}[t]
\caption{\parbox{0.9\linewidth}{Comparison of our \xname models against various baseline models. Accuracy data are cited from their original paper~\cite{liu2023chipnemo, liu2023rtlcoder, pei2024betterv, gao2024autovcodersystematicframeworkautomated, yang2025haven,cui2024origen, liu2025craftrtlhighqualitysyntheticdata}. * are the result updated by us with the new version.}}
\label{tab:main_exp}
\centering
\footnotesize
{%
\begin{tabular}{ccccccccccc}
\toprule
\multirow{2}{*}{Type} & \multirow{2}{*}{Model} & \multirow{2}{*}{\begin{tabular}[c]{@{}c@{}}Open\\ source\end{tabular}} & \multicolumn{3}{c}{VerilogEval-Machine (\%)} & \multicolumn{3}{c}{VerilogEval-Human (\%)} & \multicolumn{2}{c}{RTLLM v1.1 (\%)} \\ 
 &  & &   pass@1 & pass@5 & pass@10 & pass@1 & pass@5 & pass@10 &  Syntax & Func. \\
 \midrule
\multirow{7}{*}{\begin{tabular}[c]{@{}c@{}}Foundation models\end{tabular}}& GPT-3.5 & \text{\texttimes} & 46.7 & 69.1 & \multicolumn{1}{c}{74.1} & 26.7 & 45.8 & 51.7 & \multicolumn{1}{c}{89.7} & 37.9 \\
 & GPT-3.5-turbo-0125* & \text{\texttimes} & 60.9 & 75.0 & \multicolumn{1}{c}{79.9} & 33.5 & 45.9 & 50.0 & \multicolumn{1}{c}{79.3} & 51.7 \\
 & GPT-4 & \text{\texttimes} & 60.0 & 70.6 & \multicolumn{1}{c}{73.5} & 43.5 & 55.8 & 58.9 & \multicolumn{1}{c}{\textbf{100.0}} & \underline{65.5} \\
 & StarCoder & \checkmark & 46.8 & 54.5 & \multicolumn{1}{c}{59.6} & 18.1 & 26.1 & 30.4 & \multicolumn{1}{c}{93.1} & 27.6 \\
 & CodeLlama & \checkmark & 43.1 & 47.1 & {47.7} & 18.2 & 22.7 & 24.3 & {86.2} & 31.0 \\
 & DeepSeek-Coder & \checkmark & 52.2 & 55.4 & {56.8} & 30.2 & 33.9 & 34.9 & \multicolumn{1}{c}{93.1} & 44.8 \\
 & CodeQwen & \checkmark & 46.5 & 54.9 & \multicolumn{1}{c}{56.4} & 22.5 & 26.1 & 28.0 & \multicolumn{1}{c}{86.2} & 41.4 \\
 & Qwen2.5-Coder & \checkmark & 66.2 & 79.2  & 83.9  & 34.6  & 45.6  & 51.0  & 89.6 & 41.4 \\
 \midrule
\multirow{9}{*}{\begin{tabular}[c]{@{}c@{}}IT baselines\end{tabular}}& ChipNeMo & \text{\texttimes} & 43.4 & - & \multicolumn{1}{c}{-} & 22.4 & - & - & \multicolumn{1}{c}{-} & - \\
 & Thakur et al. & \checkmark & 44.0 & 52.6 & {59.2} & 30.3 & 43.9 & 49.6 & {86.2} & 24.1 \\
 & RTLCoder-Mistral & \checkmark & 62.5 & 72.2 & \multicolumn{1}{c}{76.6} & 36.7 & 45.5 & 49.2 & \multicolumn{1}{c}{96.6} & 48.3 \\
& RTLCoder-DS & \checkmark & 61.2 & 76.5 & {81.8} & 41.6 & 50.1 & 53.4 & {93.1} & 48.3 \\
& BetterV-CL & \text{\texttimes} & 64.2 & 75.4 & \multicolumn{1}{c}{79.1} & 40.9 & 50.0 & 53.3 & \multicolumn{1}{c}{-} & - \\
& BetterV-DS & \text{\texttimes} & 67.8 & 79.1 & \multicolumn{1}{c}{84.0} & 45.9 & 53.3 & 57.6 & \multicolumn{1}{c}{-} & - \\
& BetterV-CQ & \text{\texttimes} & 68.1 & 79.4 & {84.5} & 46.1 & 53.7 & 58.2 & \multicolumn{1}{c}{-} & - \\
& CraftRTL-CL & \text{\texttimes} & 78.1 & 85.5 & 87.8 & \underline{63.1} & 67.8 & 69.7 & 93.9 & 52.9 \\
& CraftRTL-DS & \text{\texttimes} & 77.8 & 85.5 & 88.1 & \textbf{65.4} & \textbf{70.0} & \textbf{72.1} & 84.3 & 58.8 \\
 \midrule
\multirow{4}{*}{CodeV-Verilog}& CodeV-Verilog-CL(Ours) & \checkmark & {78.1} & 86.0 & \multicolumn{1}{c}{{88.5}} & 45.2 & 59.5 & 63.8 & \multicolumn{1}{c}{93.1} & \textbf{62.1} \\
& CodeV-Verilog-DS(Ours) & \checkmark & {77.9} & \underline{88.6} & \multicolumn{1}{c}{{90.7}} & {52.7} & {62.5} & {67.3} & \multicolumn{1}{c}{89.7} & 55.2 \\
 & CodeV-Verilog-CQ(Ours) & \checkmark & 77.6 & {88.2} & \multicolumn{1}{c}{{90.7}} & {53.2} & {65.1} & {68.5} & \multicolumn{1}{c}{93.1} & 55.2 \\
  & CodeV-Verilog-QC(Ours) & \checkmark & \underline{80.1} & 87.9  & 90.5  & {59.2}  & {65.8}  & {69.1}  & \underline{96.6} & 51.7 \\
  \midrule
\multirow{4}{*}{CodeV-All}& CodeV-All-CL(Ours)  & \checkmark & 78.5 & 85.6  & 87.6  & 46.6  & 58.8  & 62.5  & {96.6} & 55.2 \\
& CodeV-All-DS(Ours) & \checkmark & 79.8 & 86.0  & 86.7  & 53.0  & 63.3  & 67.2  & {96.6} & 51.7 \\
& CodeV-All-CQ(Ours)  & \checkmark & 79.9 & 88.3  & \underline{91.1}  & 54.1  & 65.1  & 68.6  & 93.1 & 58.6 \\
& CodeV-All-QC(Ours)  & \checkmark & \textbf{81.9} & \textbf{89.9}  & \textbf{92.0}  & {56.6}  & \underline{67.9}  & \underline{71.4}  & {96.6} & 55.2 \\
 \bottomrule
\end{tabular}%
}

\end{table*}

\subsection{Experimental Setup}
\subsubsection{Supervised Fine-tuning}
We use CodeLlama-7b-Instruct \cite{codellama}, Deepseek-Coder-6.7b-Instruct~\cite{deepseekcoder}, CodeQwen1.5-7B-Chat~\cite{qwen} and Qwen2.5-Coder-7B~\cite{hui2024qwen2} as our base models. We fine-tune these models with full parameters for $3$ epochs using PyTorch's Distributed Data Parallel (DDP) on 4 NVIDIA A100-80GB SMX GPUs.
We employ the Adafactor optimizer~\cite{shazeer2018adafactor} with a linear learning rate scheduler, setting the initial learning rate at $5e-5$ with $15$ warm-up steps. We set global batch size $256$, with max sequence length $2048$ which follows the setting of previous work~\cite{liu2023rtlcoder, chang2024data}.
\subsubsection{Benchmarks and Metrics}
\label{sec:benchmarks}

We conduct comprehensive evaluations on the VerilogEval~\cite{verilogeval}, RTLLM~\cite{rtllm}, ChiselTutorial, ChiselEval, VerilogEval-FIM, RTLLM-FIM and ChiselEval-FIM benchmarks. 

VerilogEval includes two parts: VerilogEval-Machine and VerilogEval-Human, with $143$ GPT-generated and $156$ hand-crafted Verilog generation problems, respectively. We follow the VerilogEval paper, using the pass@k metric to measure the Verilog generation accuracy. The pass@k metric estimates the proportion of problems that can be solved at least once in $k$ attempts:
\begin{equation}
    \mathbf{pass}  @ k:=\underset{\text { problems }}{\mathbb{E}}\left[\frac{1-\binom{n-c}{k}}{\binom{n}{k}}\right]
    \label{eq:passk}
\end{equation}
where $n \ge k$ represents the total number of trials for each problem, and $c$ represents the number of trials that pass the functional check. We set $n = 20$ following previous work.

RTLLM contains $29$ Verilog generation problems. Following prior work~\cite{liu2023rtlcoder}, we test models on RTLLMv1.1 which is error-fixed. 
Also, following the original settings, we generate Verilog code for each problem $5$ times and report the success rate (proportion of problems that at least one trial passes the syntax check and function check, respectively). 
In addition, we also provide a more comprehensive evaluation of the RTLLM benchmark in Section~\ref{sec:rtllm_sample20}, setting $n=20$ for each generation task which is similar to VerilogEval.

Following RTLCoder~\cite{liu2023rtlcoder}, for all tests, we set the generation temperature of each model to \{0.2, 0.5, 0.8\}, and report the best performance across different temperature settings, but we also present results on three temperatures separately in Section~\ref{sec:temperature}.

Details on ChiselTutorial, ChiselEval, VerilogEval-FIM, RTLLM-FIM, and ChiselEval-FIM can refer to Section~\ref{exp:Chat_benchmark} and~\ref{exp:FIM_benchmark}. We evaluate model performance using the pass@k metric with $n=20$, except for RTLLM-FIM, which follows the same evaluation metric as RTLLM (perform code infilling for each problem $5$ times and report the success rate). We fix the temperature at 0.2 to compare the results of different models.

In all the results, We use \textbf{bold} to denote the best model overall and \underline{underline} to indicate the second-best results overall.

\subsubsection{Baseline Methods}

In our experiment, we compare our results with baseline methods on $\sim$7B LLMs, including: 
(1) Foundation models not tailored for Verilog generation tasks, including closed-source commercial LLMs such as GPT-3.5 and GPT-4, and four open-source models specialized in code generation tasks: StarCoder~\cite{starcoder}, CodeLlama-7b-Instruct~\cite{codellama}, DeepSeek-Coder-6.7b-Instruct~\cite{deepseekcoder}, CodeQwen1.5-7B-Chat~\cite{qwen}, and Qwen2.5-Coder-7B~\cite{hui2024qwen2}. 
(2) Models instruction-tuned for Verilog generation tasks (denoted as IT baselines), including ChipNeMo~\cite{liu2023chipnemo}, Thakur et al.~\cite{thakur2023benchmarking}, RTLCoder~\cite{liu2023rtlcoder}, BetterV~\cite{pei2024betterv} and CraftRTL~\cite{liu2025craftrtlhighqualitysyntheticdata}.

LLMs trained based on CodeLlama-7b-Instruct, DeepSeek-Coder-6.7b, CodeQwen1.5-7B-Chat, and Qwen2.5-Coder-7B will be denoted with the suffix CL, DS, CQ, QC, respectively.

Due to OriGen~\cite{cui2024origen}, AutoVCoder~\cite{gao2024autovcodersystematicframeworkautomated}, and HAVEN~\cite{yang2025haven} using techniques such as Self-Reflection, RAG, and CoT prompting models to enhance the performance of foundation models, which cannot be fairly compared with ours, so we have omitted the comparison with them.

\begin{table*}[t]
\centering
\caption{Comparison under same training set size. 
}
\label{tab:27k_cmp}
{
\begin{tabular}{ccccccccc}
\toprule
\multirow{2}{*}{Model} & \multicolumn{3}{c}{VerilogEval-Machine (\%)} & \multicolumn{3}{c}{VerilogEval-Human (\%)} & \multicolumn{2}{c}{RTLLM v1.1 (\%)} \\
 & pass@1 & pass@5 & pass@10 & pass@1 & pass@5 & pass@10 & Syntax & Func. \\ 
\midrule
RTLCoder-DeepSeek & \underline{61.2} & \underline{76.5} & \underline{81.8} & \textbf{41.6} & \underline{50.1} & \underline{53.4} & \textbf{93.1} & \textbf{48.3} \\
RTLCoder-DeepSeek-Direct & 59.8 & 73.6 & 77.2 & \underline{39.1} & 48.3 & 51.3 & 86.2 & \underline{44.8} \\ 
\midrule
CodeV-Verilog-DS-27K & \textbf{72.3} & \textbf{82.2} & \textbf{85.6} & \underline{39.1} & \textbf{53.4} & \textbf{58.3} & \underline{89.6} & \underline{44.8} \\ 
\bottomrule
\end{tabular}%
}

\end{table*}

\subsection{Results on VerilogEval and RTLLM}
\label{exp:main}
In this section, we compare our method with previous work fairly on VerilogEval and RTLLM.
Table~\ref{tab:main_exp} compares the main results of our \xname with baseline methods on the VerilogEval and RTLLM benchmarks. 
We evaluate CodeLlama, DeepSeek-Coder, CodeQwen and Qwen2.5-Coder , while other baseline results are sourced from RTLCoder~\cite{liu2023rtlcoder}, BetterV~\cite{pei2024betterv}, AutoVCoder~\cite{gao2024autovcodersystematicframeworkautomated}, OriGen~\cite{cui2024origen}, CraftRTL~\cite{liu2025craftrtlhighqualitysyntheticdata} and HAVEN~\cite{yang2025haven}. 
The results show that:

\textbf{\xname-Verilog performs well and achieves SOTA on Verilog-Machine.} 
\xname-Verilog performed well in the VerilogEval and RTLLM benchmarks, surpassing the RTLCoder and BetterV methods in the VerilogEval benchmark test,
particularly in the most challenging pass@1 metric where \xname-Verilog-QC achieves 80.1\% in VerilogEval-machine, and \xname-Verilog-QC achieves 59.2\% in VerilogEval-human, significantly outperforming both GPT-4 and BetterV-CodeQwen. In the RTLLM v1.1 benchmark, \xname-Verilog-CL nearly matches GPT-4's functional check success rate, failing in just 1 more case out of 29 circuits compared to GPT-4, and significantly outperforms all other models. Also, \xname-Verilog significantly outperforms its GPT-3.5-turbo-0125 summarization model. These results demonstrate that contributed by a high-quality instruction tuning dataset, \xname-Verilog exhibits significant superiority in Verilog generation tasks.

\textbf{The performance of different base models becomes closer after fine-tuning.}
 This is reasonable because the base models are trained on different training data but use the same fine-tuning data, which reduces their differences. Additionally, we find that after fine-tuning, CodeQwen surpasses DeepSeek-Coder, and CodeLlama also outperformed the other two models on VerilogEval-Machine pass@1 after fine-tuning. This indicates that the performance of a base model in specific domains does not necessarily align with its performance after fine-tuning.

\textbf{Under the same dataset size, our method yields better data.}
 As shown in Table~\ref{tab:27k_cmp}, under the same data size (\ie 27K), the results obtained from our approach surpass those of RTLCoder, especially for its directly trained version. Please note that in addition to data distillation, RTLCoder also proposes a training method based on syntax checking, which is orthogonal with our method and may improve \xname-Verilog's performance in the future. More analytical results on the impact of data size are shown in Section~\ref{sec:data size}.

\textbf{There is still a large gap between Human and Machine on VerilogEval.}
This stays the same as prior work. 
There are two reasons for this: First, the descriptions for Machine are generated by GPT, so they are closer to the distribution of LLMs compared with Human. 
Second, some of the Human descriptions contain tabular data like truth tables, which are difficult to obtain from the open-source codebase or GPT, leading to a distribution difference in the descriptions.

\begin{table}[!]
\caption{Performance of models on ChiselEval and ChiselTutorial.}
\centering
\setlength{\tabcolsep}{1mm}
\footnotesize
\small
\begin{tabular}{ccccccc}
\toprule
\multirow{2}{*}{Model} & \multicolumn{3}{c}{ChiselTutorial (\%)} & \multicolumn{3}{c}{ChiselEval (\%)} \\
  & p@1 & p@5 & p@10 & p@1 & p@5 & p@10 \\ \midrule

\multirow{1}{*}{GPT-3.5-turbo-0125} & {73.2} & {77.9} & {79.4} & {28.6} & {30.5} & {31.5} \\
\multirow{1}{*}{GPT-4} & \textbf{79.1} & \textbf{89.5} & \textbf{91.2} & \textbf{38.9} & {46.5} & {49.7} \\

\midrule
\multirow{1}{*}{CodeLlama} & {31.2} & {56.7} & {63.2} & {17.7} & {26.2} & {29.4} \\
\multirow{1}{*}{DeepSeekCoder} & {33.8} & {53.0} & {58.0} & {3.0} & {5.6} & {6.9} \\
\multirow{1}{*}{CodeQwen1.5} & {69.4} & {80.3} & {82.2} & {25.3} & {34.3} & {36.8} \\
\multirow{1}{*}{Qwen2.5-Coder} & {53.5} & {79.3} & {83.9} & {19.3} & {34.8} & {40.2} \\
\midrule

\multirow{1}{*}{RTLCoder-Mistral} & {0.3} & {1.4} & {2.9} & {0.0} & {0.0} & {0.0} \\
\multirow{1}{*}{RTLCoder-DS} & {30.5} & {37.7} & {41.1} & {1.6} & {3.3} & {4.1} \\
\midrule
\multirow{1}{*}{CodeV-All-CL} & {76.2} & {80.0} & {81.7} & {34.3} & {40.6} & {42.7} \\
\multirow{1}{*}{CodeV-All-DS} & {66.8} & {81.3} & {85.0} & {32.6} & {41.2} & {43.5} \\
\multirow{1}{*}{CodeV-All-CQ} & {78.8} & {88.2} & {88.2} & {31.1} & {40.9} & {43.7} \\
\multirow{1}{*}{CodeV-All-QC} & {74.7} & {86.6} & {88.1} & {34.4} & \textbf{47.3} & \textbf{51.0} \\
\bottomrule
\end{tabular}%
\label{tab:test_chisel-chat}
\end{table}

\begin{table*}[h]
\caption{Comparison between our \xname-All models and the base models on ChiselEval-FIM.}
\label{tab:FIM_chisel}
\centering
\footnotesize
\setlength{\tabcolsep}{1mm}
{%
\begin{tabular}{cccccccccc}
\toprule
 \multirow{3}{*}{Model}& \multicolumn{9}{c}{ChiselEval-FIM (\%)} \\
  & \multicolumn{3}{c}{single-line}&    \multicolumn{3}{c}{multi-line}&\multicolumn{3}{c}{random-span} \\ 
   &   pass@1& pass@5& pass@10&    pass@1&pass@5&pass@10&pass@1& pass@5&pass@10\\
  \midrule
  \multirow{1}{*}{CodeLlama} & 72.0 & 82.6 & \multicolumn{1}{c}{85.0 } &    36.4 &42.7 &44.4 &33.3 & 37.9 & 39.5 
\\
  \multirow{1}{*}{DeepSeekCoder} & 68.9 & 76.7 & \multicolumn{1}{c}{78.7 } &    35.0 &40.0 &41.2 &40.6 & 43.9 & 44.6 
\\
  \multirow{1}{*}{CodeQwen1.5} & 79.8 & 86.7 & \multicolumn{1}{c}{88.1 } &    47.2 &51.5 &52.3 &42.6 & 48.3 & 50.0 
\\
  \multirow{1}{*}{Qwen2.5-Coder} & 83.4 & 88.3 &  89.4 &    47.6 &52.7 &54.1 &45.9 & 50.3 &52.0 
\\
 \midrule
 \multirow{1}{*}{CodeV-All-CL} 
& 88.3 & \underline{90.4} & \multicolumn{1}{c}{90.9 } &    45.8 &48.1 &48.9 &48.2 & 51.8 &  

52.7 
\\
 \multirow{1}{*}{CodeV-All-DS} 
& \underline{89.6} & \textbf{92.3} & \multicolumn{1}{c}{\textbf{92.8} } &    \textbf{52.1} &\textbf{55.8} &\textbf{56.5} &\textbf{52.6} & \textbf{56.5} & \textbf{57.8} 
\\
  \multirow{1}{*}{CodeV-All-CQ} 
& \textbf{89.7} & \textbf{92.3} &  \underline{92.6} &    47.5 &\underline{53.8} &\underline{55.4} &49.1 & 54.1 &56.3 
\\
  \multirow{1}{*}{CodeV-All-QC} & 87.3 & \underline{90.4} & \multicolumn{1}{c}{91.3 } &    \underline{49.7} &53.3 &54.3 &\underline{51.6} & \underline{55.5} & \underline{56.6} 
\\
 \bottomrule
\end{tabular}%
}
\end{table*}

\begin{table*}[h]
\caption{Comparison between our \xname-All models and the base models on VerilogEval-FIM.}
\label{tab:FIM_Verilog}
\centering
\footnotesize
\setlength{\tabcolsep}{1mm}
{%
\begin{tabular}{ccccccccccccccccccc}
\toprule
 \multirow{3}{*}{Model}& \multicolumn{9}{c}{VerilogEval-FIM-Machine (\%)} & \multicolumn{9}{c}{VerilogEval-FIM-Human (\%)}\\
  & \multicolumn{3}{c}{single-line}&    \multicolumn{3}{c}{multi-line}&\multicolumn{3}{c}{random-span} & \multicolumn{3}{c}{single-line}& \multicolumn{3}{c}{multi-line}& \multicolumn{3}{c}{random-span}\\ 
   &   p@1& p@5& p@10&    p@1&p@5&p@10&p@1& p@5&p@10& p@1& p@5& p@10& p@1& p@5& p@10& p@1& p@5&p@10\\
  \midrule
  \multirow{1}{*}{CodeLlama} & 57.6 & 65.2 & \multicolumn{1}{c}{67.3 } &    40.0 &48.5 &51.2 &42.2 & 50.4 & 52.5 
& 56.7 & 64.5 & 66.0 & 39.9 & 47.7 & 50.0 & 33.9 & 40.1 &41.6 
\\
  \multirow{1}{*}{DeepSeekCoder} & 65.8 & 69.5 & \multicolumn{1}{c}{70.7 } &    45.6 &48.7 &49.7 &51.6 & 55.0 & 55.7 
& 63.1 & 68.1 & 70.3 & 43.8 & 49.0 & 50.6 & 43.2 & 48.7 &50.0 
\\
  \multirow{1}{*}{CodeQwen1.5} & 59.6 & 68.4 & \multicolumn{1}{c}{71.8 } &    41.9 &48.1 &50.7 &41.9 & 48.1 & 50.2 
& 55.1 & 64.3 & 66.6 & 34.8 & 41.9 & 43.2 & 38.6 & 43.9 &45.8 
\\
  \multirow{1}{*}{Qwen2.5-Coder} & 71.2 & 74.2 &  74.6 &    48.1 &54.2 &55.4 &53.3 & 58.4 &59.9 
& 67.2 & 70.7 & 71.4 & 44.8 & 48.5 & 49.3 & 42.1 & 47.8 &49.7 
\\
 \midrule
 \multirow{1}{*}{CodeV-All-CL} & 76.9 & 78.9 & \multicolumn{1}{c}{79.6 } &    63.6 &67.0 &67.8 &61.1 & 66.4 &  

 68.0 
& \underline{75.6} & \underline{78.1} & \underline{78.6} & 55.4 & 58.7 & 60.0 & 52.2 & 56.3 &
57.6 
\\
 \multirow{1}{*}{CodeV-All-DS} & 74.3 & 80.0 & \multicolumn{1}{c}{82.2 } &    63.2 &69.5 &71.2 &65.2 & 69.2 & 71.0 
& 71.2 & 75.6 & 77.2 & 59.2 & \underline{65.0} & \underline{66.8} & \underline{58.9} & \underline{63.9} &\underline{65.5} 
\\
  \multirow{1}{*}{CodeV-All-CQ} & \underline{79.0} & \underline{82.3} &  \underline{83.3} &    \underline{65.5} &\underline{69.6} &\underline{71.3} &\underline{69.4} & \underline{71.6} &\underline{72.0} 
& 72.7 & 76.1 & 77.6 & \underline{59.5} & 62.7 & 63.9 & 58.2 & 63.3 &64.8 
\\
  \multirow{1}{*}{CodeV-All-QC} & \textbf{80.0} & \textbf{84.3} & \multicolumn{1}{c}{\textbf{85.5} } &    \textbf{72.8} &\textbf{74.8} &\textbf{75.5} &\textbf{70.5} & \textbf{73.4} & \textbf{73.9} 
& \textbf{76.4} & \textbf{79.1} & \textbf{80.0} & \textbf{62.3} & \textbf{66.5} & \textbf{67.6} & \textbf{64.7} & \textbf{68.4} &\textbf{69.3} 
\\
 \bottomrule
\end{tabular}%
}
\end{table*}

\begin{table}[!]
\caption{The performance of CodeV-All models on our Chisel benchmark.}
\label{tab:FIM_rtllm}
\centering
\setlength{\tabcolsep}{1mm}
\footnotesize
\small
{%
\begin{tabular}{ccccccc}
\toprule
 \multirow{3}{*}{Model}& \multicolumn{6}{c}{RTLLM-FIM (\%)}\\
  & \multicolumn{2}{c}{single-line}&    \multicolumn{2}{c}{multi-line}&\multicolumn{2}{c}{random-span}\\ 
   &   Syntax& Func.&    Syntax&Func.&Syntax& Func.\\
  \midrule
  \multirow{1}{*}{CodeLlama} & \underline{96.6} & 79.3 &    93.1 &48.3 &82.8 & 37.9 
\\
  \multirow{1}{*}{DeepSeekCoder} & \underline{96.6} & \underline{93.1} &    \underline{96.6} &58.6 &93.1 & 44.8 
\\
  \multirow{1}{*}{CodeQwen1.5} & 93.1 & 86.2 &    89.7 &55.2 &82.8 & 34.5 
\\
  \multirow{1}{*}{Qwen2.5-Coder} & \textbf{100.0} & \underline{93.1} &    \underline{96.6} &55.2 &\underline{96.6} & 44.8 
\\
 \midrule
 \multirow{1}{*}{CodeV-All-CL} 
& \textbf{100.0}&  89.7 &    \underline{96.6} &55.2 &\underline{96.6} & 

44.8 
\\
 \multirow{1}{*}{CodeV-All-DS} 
& \textbf{100.0} & \textbf{96.6} &    \textbf{100.0} &\textbf{72.4} &\textbf{100.0} & \textbf{62.1} 
\\
  \multirow{1}{*}{CodeV-All-CQ} 
& \textbf{100.0} & \underline{93.1} &    \textbf{100.0} &58.6 &\textbf{100.0} & 55.2 
\\
  \multirow{1}{*}{CodeV-All-QC} & \textbf{100.0} & \textbf{96.6} &    \textbf{100.0} &\underline{62.1} &93.1 & \underline{58.6} 
\\
 \bottomrule
\end{tabular}%
}
\vspace{-10px}
\end{table}

\begin{table}[!]
\caption{Compare the performance of the models trained on 165K Verilog and 184K Verilog-Chisel datasets}
\centering
\setlength{\tabcolsep}{1mm}
\footnotesize
\small
\begin{tabular}{ccccccc}
\toprule
\multirow{2}{*}{Model} & \multicolumn{3}{c}{ChiselTutorial (\%)} & \multicolumn{3}{c}{ChiselEval (\%)} \\
  & p@1 & p@5 & p@10 & p@1 & p@5 & p@10 \\ \midrule
  
\multirow{1}{*}{Qwen2.5-Coder} & {53.5} & {79.3} & {83.9} & {19.2} & {34.8} & {40.2} \\

\multirow{1}{*}{CodeV-Verilog-QC(Ours)} & {0} & {0} & {0} & {0} & {0} & {0} \\

\multirow{1}{*}{CodeV-All-QC(Ours)}  & \textbf{74.7} & \textbf{86.6} & \textbf{88.1} & \textbf{34.4} & \textbf{47.3} & \textbf{51.0} \\

\bottomrule
\end{tabular}%
\label{tab:chisel_verilog}
\end{table}

\subsection{Results on Multi-lingual and Multi-Scenario Tasks}
\label{exp:codev-chisel-fim}
We evaluate LLMs' capabilities for multi-lingual and multi-scenario tasks in this section.
As mentioned before, we added 18.7K Chisel data extension based on the existing 165K Verilog dataset, thereby constructing a more comprehensive training corpus, 184K-Verilog-Chisel. 
Then, we utilize the Chat-FIM-Tag supervised fine-tuning process to develop the \xname-All series, which can handle tasks in different languages (Verilog and Chisel) and scenarios (Chat and FIM).

To evaluate the models' multi-lingual (Verilog and Chisel) generation capabilities on Chat tasks, we evaluate both our CodeV-All and baseline LLMs across four benchmarks: VerilogEval, RTLLM, ChiselTutorial, and ChiselEval. The results are presented in Tables~\ref{tab:main_exp} and \ref{tab:test_chisel-chat}. To evaluate the models' multi-scenario ability, we evaluate CodeV-All on the proposed FIM benchmarks. The results are shown in Tables~\ref{tab:FIM_chisel}, \ref{tab:FIM_Verilog}, and \ref{tab:FIM_rtllm}. We can conclude that:

\begin{table}[!]
\caption{Ablation study on multi-level summarization (MLS) on VerilogEval-Machine and VerilogEval-Human with pass@k (p@k). 
}
\centering
\setlength{\tabcolsep}{1mm}
\footnotesize
\small
\begin{tabular}{cccccccc}
\toprule
\multirow{2}{*}{Model} & \multirow{2}{*}{Temp.} & \multicolumn{3}{c}{VE-Machine (\%)} & \multicolumn{3}{c}{VE-Human (\%)} \\
 &  & p@1 & p@5 & p@10 & p@1 & p@5 & p@10 \\ \midrule
\multirow{3}{*}{\begin{tabular}[c]{@{}c@{}}CodeV-Verilog-CL\\ w/o MLS\end{tabular}} & 0.2 & 59.0 & 66.3 & \multicolumn{1}{c}{68.9} & 28.3 & 37.5 & 40.3 \\
 & 0.5 & 59.5 & 72.1 & \multicolumn{1}{c}{75.8} & 25.1 & 40.4 & \underline{45.7} \\
 & 0.8 & 54.7 & 72.7 & \multicolumn{1}{c}{\underline{77.5}} & 21.4 & 39.2 & 45.0 \\ \midrule
\multirow{3}{*}{\begin{tabular}[c]{@{}c@{}}CodeV-Verilog-CL\\  w/ MLS\end{tabular}} & 0.2 & \textbf{63.0} & 71.2 & \multicolumn{1}{c}{72.9} & \textbf{31.7} & 38.1 & 39.8 \\
 & 0.5 & \underline{60.9} & \underline{73.1} & \multicolumn{1}{c}{76.5} & \underline{29.2} & \underline{41.4} & \underline{45.7} \\
 & 0.8 & 58.5 & \textbf{73.9} & \multicolumn{1}{c}{\textbf{78.1}} & 25.7 & \textbf{41.5} & \textbf{47.6} \\
\bottomrule
\end{tabular}%

\label{tab:ablation}
\vspace{-10px}
\end{table}

\begin{table*}[!]
\caption{The ablation study of tag on the 20k data.}
\centering
\setlength{\tabcolsep}{1mm} %
\begin{tabular}{ccccccccccccc}
\toprule
\multirow{2}{*}{Model} & \multicolumn{3}{c}{ChiselTutorial (\%)} & \multicolumn{3}{c}{ChiselEval (\%)} & \multicolumn{3}{c}{VerilogEval-Machine (\%)} &  \multicolumn{3}{c}{VerilogEval-Human (\%)} \\
  & pass@1 & pass@5 & pass@10 & pass@1 & pass@5 & pass@10 & pass@1 & pass@5 & pass@10 & pass@1 & pass@5 & pass@10 \\ \midrule
\multirow{2}{*}{\parbox{2.5cm}{\centering CodeV-All-QC-20K\\ w/o Tag}} & 
\multirow{2}{*}{70.0} & \multirow{2}{*}{86.4} & \multirow{2}{*}{88.1} & 
\multirow{2}{*}{29.2} & \multirow{2}{*}{39.6} & \multirow{2}{*}{44.1} & \multirow{2}{*}{67.2} & \multirow{2}{*}{75.5} & \multirow{2}{*}{78.2} & \multirow{2}{*}{42.9} & \multirow{2}{*}{54.2} & \multirow{2}{*}{59.1} \\
& & & & & & \\
\multirow{2}{*}{\parbox{2.5cm}{\centering CodeV-All-QC-20K\\ w/ Tag}} & 
\multirow{2}{*}{\textbf{78.8}} & \multirow{2}{*}{\textbf{91.0}} & \multirow{2}{*}{\textbf{93.4}} & 
\multirow{2}{*}{\textbf{30.5}} & \multirow{2}{*}{\textbf{40.8}} & \multirow{2}{*}{\textbf{44.6}}  & \multirow{2}{*}{\textbf{69.9}} & \multirow{2}{*}{\textbf{77.0}} & \multirow{2}{*}{\textbf{80.2}} & \multirow{2}{*}{\textbf{46.3}} & \multirow{2}{*}{\textbf{56.8}} & \multirow{2}{*}{\textbf{60.7}}\\
& & & & & & \\
\bottomrule
\end{tabular}
\label{tab:ablation_chisel_verilog}
\vspace{-10px}
\end{table*}
\textbf{The inclusion of the Chisel dataset leads to an improvement in Chisel performance.}
To verify the efficiency of our method, we compare CodeV-Verilog and CodeV-All on ChiselTutorial and ChiselEval in Table~\ref{tab:chisel_verilog}. Compared to the data using only Verilog, our model significantly enhanced the Qwen2.5Coder's ability on Chisel, achieving a pass@1 of 74.7\%, pass@5 of 86.6\%, and pass@10 of 88.1\% in the ChiselTutorial test, and a pass@1 of 34.4\%, pass@5 of 47.3\%, and pass@10 of 51.0\% in the Chisel-Eval test.

\textbf{\xname-All demonstrates strong performance on Chisel generation tasks.}
\xname-All-CQ achieves 78.8\% on pass@1, 88.2\% on pass@5, and 88.2\% on pass@10 on the ChiselTurtrial benchmark. On the ChiselEval benchmark, \xname-All-QC achieves 34.4\% on pass@1, 47.3\% on pass@5, and 51.0\% on pass@10. 

\textbf{\xname-All performs comparable to \xname-Verilog on Verilog tasks.}
\xname-All maintains high Verilog generation quality and grammatical correctness as shown in Table~\ref{tab:main_exp}. This finding suggests that the newly added Chisel data and FIM fine-tuning strategy don't negatively impact the model's original Verilog generation capability, but rather effectively expand its functionality.

\textbf{Our fine-tuning method is effective in improving the model's infilling performance.}
Due to the inconsistent and non-standardized infilling outputs generated by the base models in FIM tasks (all base models show poor performance), we apply post-processing to the outputs of the base models. That is, we identify and remove redundant modules, truncate the outputs at special tokens, and individually address certain unexpected outputs generated by the base models. It is important to note that \xname-All does not engage in any post-processing beyond the removal of a redundant space character that may appear at the beginning of the output. 
The results show that \xname-All achieves significant improvements across all three FIM benchmarks following the infilling fine-tuning process. This highlights the effectiveness of our method in enhancing the model's infilling capabilities, even when facing multi-lingual scenarios. 
Specifically, compared with the base models, \xname-All achieves average performance improvements of 17.0\% on VerilogEval-FIM-Machine and 13.7\% on VerilogEval-FIM-Human (Table~\ref{tab:FIM_Verilog}). On RTLLM-FIM, \xname-All surpasses the base models by an average of 5.7\% in syntax accuracy and 9.5\% in functional correctness (Table~\ref{tab:FIM_rtllm}). Additionally, an improvement of 8.3\% is observed on ChiselEval-FIM (Table~\ref{tab:FIM_chisel}).

\subsection{Ablation Study}
\label{exp:ablation}

In this section, we will measure the effectiveness of multi-level summarization and adding language tags in fine-tuning. Due to budget constraints, we conduct this experiment only on a 10K Verilog dataset. We sample 10K Verilog data from the multi-level summarization dataset and generate single-level summarization (\ie summarize directly) for the same Verilog code. The results in Table~\ref{tab:ablation} show that the instruction data obtained through multi-level summarization can effectively enhance model performance which is consistent with our expectations.

\label{exp:tag_analysis}
To verify that adding tags helps LLMs in low-resource language, we train comparative LLMs on a mixed dataset of 20K Verilog and Chisel codes. The results, as shown in Table~\ref{tab:ablation_chisel_verilog}, demonstrate that adding tags can indeed improve LLMs' performance on both Chisel and Verilog.

\begin{table}[t]
\caption{Evaluations on RTLLM using pass@k metrics ($n=20$). 
}
\centering
\footnotesize
\setlength{\tabcolsep}{1mm}
{
\begin{tabular}{cccccccc}

\toprule
\multirow{2}{*}{Model} & \multicolumn{3}{c}{RTLLM v1.1 Syntax (\%)} & \multicolumn{3}{c}{RTLLM v1.1 Func. (\%)} \\
& p@1 & p@5 & p@10 & p@1 & p@5 & p@10 \\
\midrule
 CodeLlama & 59.4 & 81.1 & 86.7 & 16.7 & 29.9 & 36.6 \\
 DeepSeek-Coder & 75.6 & 89.0 & 92.2 & 29.8 & 40.1 & 44.8 \\
 CodeQwen & 72.0 & 83.5 & 85.7 & 32.4 & 41.6 & 44.6 \\
\midrule

 RTLCoder-Mistral & 66.8 & 76.2 & 81.0 & 25.3 & 38.6 & 43.8 \\
 RTLCoder-DS & 75.9 & 86.8 & 89.2 & 37.0 & 41.7 & 44.6 \\
\midrule

 CodeV-Verilog-CL & {81.7} & {92.3} & {93.0} & {40.8} & 52.0 & 54.9 \\
 CodeV-Verilog-DS & 81.0 & 90.4 & 92.2 & \textbf{43.9} & \underline{53.3} & {55.0} \\
 CodeV-Verilog-CQ & \underline{81.5} & \underline{92.6} & \underline{95.6} & 37.9 & \textbf{55.1} & \textbf{63.4} \\
 CodeV-Verilog-QC & \textbf{91.0} & \textbf{96.4} & \textbf{96.5} & \underline{42.4} & {51.8} & \underline{57.4} \\
\bottomrule
\end{tabular}%
}
\label{tab:rtllm_sample20}
\end{table}

\subsection{Further Experimental Analysis}
\label{exp:further_analysis}
\subsubsection{Impact of Data Scaling}
\label{sec:data size}
\begin{figure}[t]
    \centering
    \includegraphics[width=0.96\linewidth]{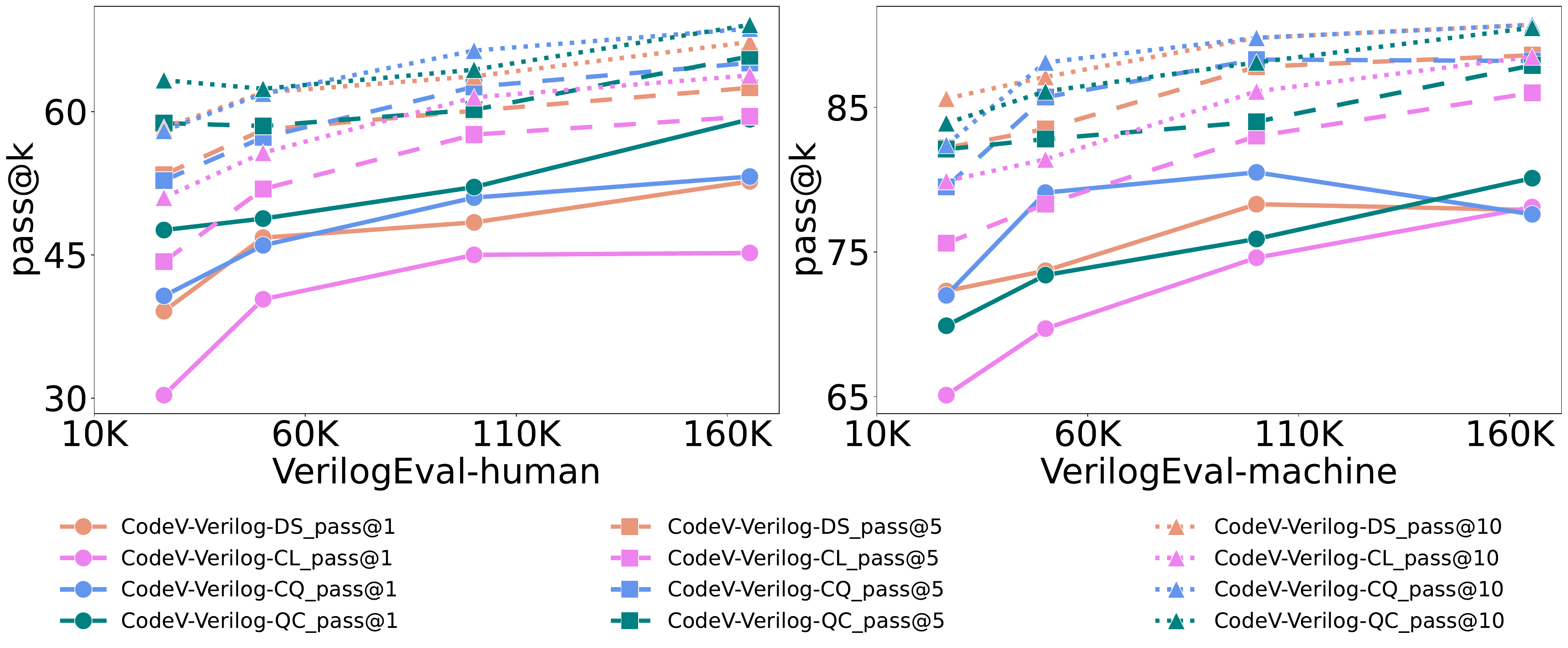}
    \caption{Analysis on dataset sizes. We train \xname-Verilog on different-size datasets and evaluate them on VerilogEval. The performance of models improves with the increase in training set size.}
    \label{fig:training_set_size}
\end{figure}
\begin{figure}[t]
    \centering
    \includegraphics[width=0.96\linewidth]{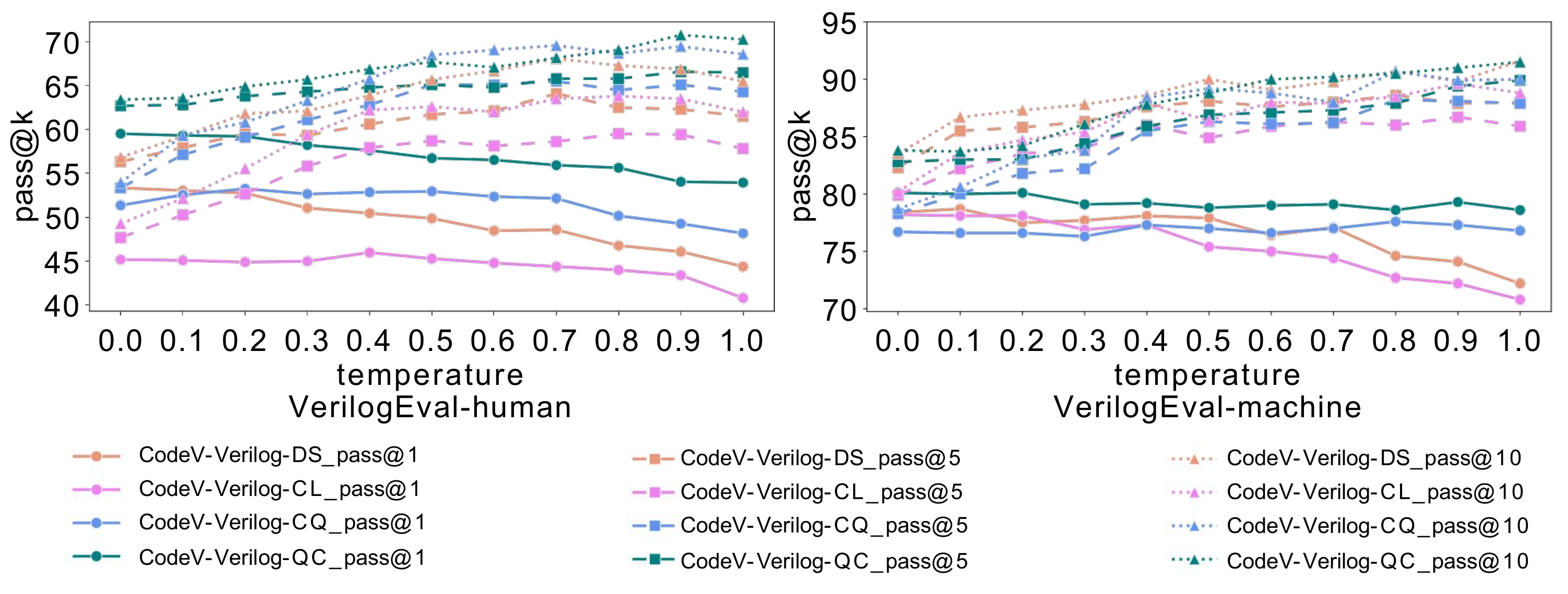}
    \caption{Analysis on generation temperature. We test our models on VerilogEval, setting different temperatures. Generally, as the temperature rises, the pass@1 metric decreases while pass@5 and pass@10 increase.}
    \label{fig:temperature}
\end{figure}
We randomly sample subsets of various sizes (27K, 50K, 100K) from the complete 165K Verilog dataset and fine-tune all three baseline models, reporting their evaluation results on VerilogEval as shown in Figure~\ref{fig:training_set_size}. Except for \xname-Verilog-CQ on VerilogEval-machine, which peaks on 100K-size, we observed that model performance generally improves as training set size increases, demonstrating the scalability of our methods.

\subsubsection{Impact of Sampling Temperature}
\label{sec:temperature}

Figure~\ref{fig:temperature} further analyzes the VerilogEval results of our models when setting on different sampling temperatures.
Generally, as the temperature increases, the pass@1 metric decreases while pass@5 and pass@10 increase.
This is because when the temperature rises, the model tends to generate more uncertain and diverse responses, which has two effects: on the one hand, uncertain responses are more likely to produce incorrect results, lowering the accuracy of individual Verilog codes.
On the other hand, the diversity enhances the exploration of various potential responses, thereby increasing the probability of at least one correct result being present among multiple Verilog codes.

\subsubsection{A More Robust Evaluation on RTLLM with pass@k Metric}
\label{sec:rtllm_sample20}
As shown in Eq.~\ref{eq:passk}, the pass@k metric provides unbiased estimations. In this section, we set $n=20$ to further evaluate our models on RTTLM using pass@k metrics and conduct varying comparisons with baselines. 
Table~\ref{tab:rtllm_sample20} illustrates that \xname-Verilog performs better than all other baseline models, especially for functional correctness.
Please note that there is a discrepancy in the pass@5 metric between Table~\ref{tab:rtllm_sample20} and Table~\ref{tab:main_exp}. This difference is due to the varying sample sizes used in the two tables. We have thoroughly reviewed the results to ensure there are no errors.

\section{Conclusion}
In this paper, we propose a fine-tuning pipeline, a series of open-source LLMs \xname, a collection of open-source datasets, and several new benchmarks for multilingual (Verilog and Chisel) HDL generation in various scenarios (Chat and FIM). Our experimental results have shown \xname's effectiveness with their superior performance over both previous open-source and commercial LLMs on Verilog and Chisel.
This highlights its potential utility in the domain of processor design automation and we believe our work represents a significant step forward in the application of LLMs to HDL generation and will serve as a valuable resource for both academia and industry.

\bibliography{references}

\vfill

\end{document}